\documentclass[12pt]{article}

\pdfoutput=1
\usepackage[latin9]{inputenc}
\usepackage[top=80pt,bottom=85pt,left=67pt,right=67pt]{geometry}
\usepackage{amssymb,amsmath}
\usepackage{xypic,subfigure}
\usepackage{graphicx,color}
\usepackage{bbm}
\usepackage{float}
\usepackage{tensor}
\usepackage{cancel}
\usepackage{slashed}
\usepackage[debug,pageanchor=false]{hyperref}
\definecolor{link}{rgb}{.8,.15,.1}
\hypersetup{colorlinks=true,linkcolor=link,citecolor=link,urlcolor=link,linktocpage}
\DeclareMathOperator{\sech}{sech}

\usepackage[
  backend=biber,
  style=numeric-comp,
  sorting=none,
  sortcites=true,
  giveninits=true,
  maxbibnames=99,   
  maxcitenames=2,    
  doi=false
]{biblatex}
\hypersetup{
    colorlinks=true,
    urlcolor=blue,
}

\makeatletter
\@addtoreset{equation}{section}
\makeatother

\allowdisplaybreaks

\newcommand{\beq}{\begin{equation}}
\newcommand{\eeq}{\end{equation}}
\newcommand{\bea}{\begin{eqnarray}}
\newcommand{\eea}{\end{eqnarray}}
\newcommand{\nn}{\nonumber}

\newcommand{\eq}{\begin{equation}}
\newcommand{\feq}{\end{equation}}
\newcommand{\eqn}{\begin{eqnarray}}
\newcommand{\feqn}{\end{eqnarray}}

\usepackage{ragged2e}

\begin{document}

\begin{titlepage}

\begin{center}

\vskip .5in 
\noindent


{\fontsize{15.84}{19}\selectfont\bfseries
${\mathcal{N}}=3$ supersymmetric AdS$_4$ solutions in $d=11$ supergravity
}

\bigskip\medskip

Andrea Conti\footnote{contiandrea@uniovi.es}\\

\bigskip\medskip
{\small 

Department of Physics, University of Oviedo, \\
C/ Leopoldo Calvo Sotelo 18, 33007 Oviedo}

\medskip
{\small and}

\medskip
{\small 

Instituto Universitario de Ciencias y Tecnolog\'ias Espaciales de Asturias (ICTEA),\\
Calle de la Independencia 13, 33004 Oviedo, Spain}

\vskip 2cm 

     	{\bf Abstract }\\[2mm]
     	
\end{center}

\justifying
\noindent We derive the necessary and sufficient conditions for AdS$_4$ solutions of $d=11$ supergravity to preserve ${\mathcal{N}} \leq 3$ supersymmetry. We use our setup to first impose ${\mathcal{N}}=2$ supersymmetry, recovering a previously known class of solutions, providing a non-trivial consistency of our equations. We then focus on the ${\mathcal{N}} = 3 $ case, first deriving the Tri-Sasaki class and then proving its uniqueness as the only ${\mathcal{N}} = 3$ AdS$_4$ class in eleven dimensional supergravity. Our work proves that all supersymmetric AdS$_4$ solutions preserving ${\mathcal{N}} > 2 $ supersymmetry are necessarily of Tri-Sasaki type.
     	\noindent
    	
\noindent

\vfill
\eject

\end{titlepage}

\tableofcontents

\section{Introduction}

Starting with the seminal work of Freund and Rubin \cite{Freund:1980xh}, AdS$_4$ solutions of eleven dimensional supergravity have played a prominent role in string theory. Early interest in these solutions was largely motivated by their consistent truncations to four dimensional theories with a cosmological constant. With the advent of the AdS/CFT correspondence, these solutions acquired renewed significance as holographic duals of three dimensional SCFTs. In particular, the connection with the ABJM theory \cite{Aharony:2008ug} provided a concrete and highly influential realization of this correspondence, stimulating much study of AdS$_4$ and their
dual Chern-Simons matter theories.

A further motivation for studying AdS$_4$ solutions in M-theory comes from their interpretation in terms of wrapped M5-branes. In particular, when the four-form flux is purely magnetic, such solutions can be interpreted as describing M5-branes wrapped on supersymmetric cycles \cite{Gauntlett:2006ux,Pernici:1984nw,Acharya:2000mu,Gauntlett:2000ng}.

There has been considerable effort over the last years to classify AdS$_4$ solutions of eleven dimensional supergravity with different amounts of supersymmetry. A number of results have been obtained within the Freund-Rubin ansatz \cite{Freund:1980xh}. For solutions of the form AdS$_4 \times \text{M}_7$, preservation of minimal supersymmetry requires M$_7$ to admit a weak G$_2$-structure. The class AdS$_4\times \text{SE}^7$, where SE$^7$ denotes a seven dimensional Sasaki-Einstein manifold, generically preserves ${\cal{N}}=2$ supersymmetry, while a Tri-Sasaki structure on M$_7$ enhances the supersymmetry to ${\cal{N}}=3$. Finally, it was shown in \cite{Haupt:2017bnj} that, for ${\cal{ N}}>4$, irrespective of what flux components one allows for, all internal spaces are orbifolds of the seven-sphere, which preserves ${\cal{N}}=8$ when no orbifold is present.

Beyond the Freund-Rubin class, substantial progress has also been made in the classification and construction of AdS$_4$ solutions in both type II supergravity and M-theory \cite{Passias:2018zlm,Gabella:2012rc,DeLuca:2018buk,Passias:2017yke,Pang:2015vna,Grigorian:2017okn,Marchesano:2020qvg,Guarino:2015jca,Guarino:2020flh,Guarino:2017jly,Guarino:2020gfe,Guarino:2024zgq,Varela:2015uca,Bobev:2020qev,Bobev:2019dik,Gauntlett:2004hh,Lozano:2014ata,Lozano:2015cra,Lozano:2016wrs,DHoker:2007zhm,DHoker:2007hhe,Bobev:2020fon,Grana:2006kf,Grana:2005sn,Merrikin:2021smb,Petrini:2009ur,Aharony:2010af,Tomasiello:2007eq,Koerber:2008rx,Gaiotto:2009yz,Lust:2004ig,Lust:2009mb,Bovy:2005qq,Corrado:2001nv,Gabella:2011sg,Martelli:2008rt,Martelli:2009ga}. These works employ a variety of different approaches, including consistent truncation to gauged supergravity as well as direct type II/M-theory constructions. In this work we study solutions in $d=11$ for which the metric is a warped product of AdS$_4$ over a seven dimensional internal manifold and the fluxes are allowed to have both electric and magnetic components.

A powerful tool in the study of such solutions is provided by G-structure techniques, which have been employed in many of the classifications mentioned above, \cite{Passias:2018zlm,Gabella:2012rc,DeLuca:2018buk,Passias:2017yke}\footnote{For example, see \cite{Tomasiello:2022dwe} for a complete and detailed review of this topic.}. The idea is to reformulate the data contained in the Killing spinor equation in a set of differential constraints that determine the properties of the internal manifold, putting on equal footing the equations of motion and the supersymmetry constraints.

The focus of this article is to extend the previous classifications of AdS$_4$ solutions in eleven dimensions to the ${\cal{N}} = 3$ supersymmetric case. We employ G-structure techniques to derive necessary and sufficient conditions for supersymmetry and then solve them. In doing so, it is useful to first derive the corresponding conditions for ${\cal{N}}=1$ solutions - these have been partially investigated in \cite{Lukas:2004ip,Behrndt:2005im}, but \cite{Gabella:2012rc} pointed out that they were not fully general, like not admitting an electric component for the flux. Therefore we begin by deriving the ${\cal{N}}=1$ supersymmetry conditions independently from first principles using our G-structure formalism -- see also \cite{Grigorian:2017okn} for a perspective based on octonions. The only previously known class in eleven dimensions that supports ${\cal{N}}=3$ supersymmetry is the so-called Tri-Sasaki class of \cite{Acharya:1998db}.  In massive type IIA, \cite{DeLuca:2018buk,Pang:2015vna} constructed and studied ${\cal{N}} = 3$ preserving AdS$_4$ solutions, which of course do not have a $d=11$ uplift. A natural question is whether more general ${\cal{N}}=3$ solutions also exist in eleven dimensions. Our main result is that any AdS$_4$ solution of eleven dimensional supergravity preserving ${\cal{N}}>2$ supersymmetry necessarily has a Tri-Sasaki internal manifold.~\\

The article is organized as follows:~\\

In section \ref{sec:Neq1class} we begin by deriving the set of necessary and sufficient supersymmetry conditions for an AdS$_4$ solution in $d=11$ supergravity to preserve minimal supersymmetry, \textit{i.e.} ${\cal{N}}=1$. 

We proceed in section \ref{sec:Neq123} to consider extended supersymmetry: focusing on deriving the set of conditions to realise ${\cal{N}}=3$ supersymmetry. These contain the conditions for ${\cal{N}}=2$ supersymmetry as a sub case, by setting one of the three spinors to vanish.

In section \ref{sec:Neq2class} we test the conditions derived in section \ref{sec:Neq123}. In particular we independently derive the class of ${\cal{N}}=2$ AdS$_4$ obtained in \cite{Gabella:2012rc} and reproduce the same conditions governing the class. This provides a strong cross check that the conditions of section \ref{sec:Neq123} are indeed correct.

In section \ref{sec:Neq3class} we focus on the case where the background preserves ${\cal{N}}=3$ supersymmetry. We find a branching of supersymmetry conditions giving two cases to consider. We show that the first case is the Tri-Sasaki class, providing another independent argument that the supersymmetry conditions of section \ref{sec:Neq123} are correct. We then show that there are no solutions to the second branch, this constitutes a proof that all the AdS$_4$ solutions preserving ${\cal{N}} \geq 3$ supersymmetry are Tri-Sasaki.

We conclude with a discussion section on the future outlook. The article is supplemented with extensive technical appendices.

\section{${\cal{N}} = 1$ AdS$_4$ conditions} \label{sec:Neq1class}

In this section we present necessary and sufficient conditions for AdS$_4$ solutions in $d=11$ supergravity to preserve ${\cal{N}}=1$ supersymmetry. We summarise the main results here and refer the reader to appendix \ref{app:Neq1} for further details.
~\\

The most general ansatz for an AdS$_4$ solution of eleven dimensional supergravity is
\begin{equation}\label{eq:Ansatz11dbosonic}
ds^2 = e^{2A} ds^2(\text{AdS}_4) + ds^2(\text{M}_7), \qquad G = c \, \text{vol}(\text{AdS}_4) + G_4,
\end{equation}
where $G$ is the four-form flux. The warp factor $e^{2A}$ and the four-form $G_4$ have support only on M$_7$ while $c$ is a constant\footnote{In appendix \ref{app:Neq1} we show that supersymmetry implies $c$ is constant for any value of $m$, where $m$ is the inverse radius of AdS$_4$.}. This form guarantees that the SO($3,2$) isometry of AdS$_4$ is preserved. The flux equation of motion \eqref{eq:eom11Dflux} and the Bianchi identity \eqref{eq:Bianchi11D} decompose into independent constraints orthogonal and parallel to vol$(\text{AdS}_4)$, yielding the following $d=7$ conditions
\begin{subequations}\label{eq:fluxBianchieqs7d}
\begin{align}
& d ( e^{4A} \star_7 G_4 ) + c \, G_4 = 0, \label{eq:fluxeq7d} \\[2mm]
& d G_4 = 0, \label{eq:Bianchieq7d}
\end{align}
\end{subequations}
which hold away from localised sources, and where $\star_7$ is the Hodge dual on M$_7$. For more details on our conventions we refer the reader to appendix \ref{sec:Conventions}.
~\\

The eleven dimensional Majorana spinor should decompose in terms of the Killing spinor on AdS$_4$, whose chiral components $\zeta_{\pm}$ obey
\begin{equation}\label{eq:KSEAdS4chiralNeq1}
\nabla_{\mu}^{(4)} \zeta_{\pm} = \frac{m}{2} \gamma_{\mu}^{(4)} \zeta_{\mp},
\end{equation}
Here $\pm$ denotes the $d=4$ chirality. We take $\zeta_{\pm}$ to be a Majorana conjugate pair, $m$ is the inverse radius of AdS$_4$. The general embedding of these spinors into $d=11$ is given by
\begin{equation}\label{eq:spinor11DNeq1}
\epsilon = \zeta_+ \otimes \chi + \zeta_- \otimes \chi^c,
\end{equation}
which is Majorana by construction, where $\chi$ is a generic Dirac spinor on M$_7$. The ansatz \eqref{eq:spinor11DNeq1} preserves at least 4 of the starting 32 supercharges.
~\\

In appendix \ref{app:Neq1} we derive geometric constraints on bilinears formed from $\chi$ that imply $d=11$ supersymmetry. To achieve this we made use of the necessary and sufficient conditions for supersymmetric ${\cal{N}}=(1,0)$ AdS$_3$ solutions to exist \cite{Legramandi:2020txf}, and impose that they experience an enhancement to AdS$_4$\footnote{This can be done without loss of generality, since AdS$_4$ admits a foliation by AdS$_3$ slices over an interval.}.
Using the seven dimensional spinor $\chi$, we define the following bilinears
\begin{equation}\label{eq:defXYNeq1}
\chi^{\dagger}\chi = e^A, \qquad X_n = \frac{1}{n!} \chi^{\dagger} \gamma_{a_1...a_n} \chi e^{a_1... a_n}, \qquad Y_n = \frac{1}{n!} \chi^{c \dagger} \gamma_{a_1 ... a_n} \chi e^{a_1 ... a_n}.
\end{equation}
From this starting point, in appendix \ref{app:Neq1} we derive necessary and sufficient conditions for the background \eqref{eq:Ansatz11dbosonic} to preserve minimal supersymmetry. In summary, one can show that the necessary and sufficient conditions for ${\cal{N}}=1$ AdS$_4$ are

\begin{subequations}\label{eq:BPSAdS4Neq1}
\begin{align}
& 3 e^{2A} m \text{Im}Y_0 - c = 0, \label{eq:BPSAdS4Neq1a} \\[2mm]
& d \left( e^{2A} \text{Re}Y_0 \right) - 2 m e^{A} \text{Re}X_1 = 0, \label{eq:BPSAdS4Neq1b} \\[2mm]
& 6 e^A \star_7 dA - \text{Im}X_2 \wedge G_4 = 0, \label{eq:BPSAdS4Neq1c} \\[2mm]
& (6 m e^{4A} - 2 c \text{Im}Y_0 ) \text{vol}_7 - e^{4A} \text{Im}Y_3 \wedge G_4 = 0, \label{eq:BPSAdS4Neq1d} \\[2mm]
& 2 \text{Re}Y_0 c \text{vol}_7 + e^{4A} \text{Re}Y_3 \wedge G_4 = 0, \label{eq:BPSAdS4Neq1e} \\[2mm]
& d \left( e^{3 A} \text{Im}X_2 \right) + 3 m e^{2A} \text{Im}Y_3 - e^{4A} \star_7 G_4 = 0, \label{eq:BPSAdS4Neq1f} \\[2mm]
& d \left( e^{2 A} \text{Im}Y_3 \right) + e^{2A} \text{Im}Y_0 G_4 = 0, \label{eq:BPSAdS4Neq1g}\\[2mm]
& d \left( e^{2 A} \text{Re}Y_3 \right) - 2 m e^A \text{Re}X_4 + e^{2A} \text{Re}Y_0 G_4 = 0, \label{eq:BPSAdS4Neq1h}\\[2mm]
& d \left( e^A \text{Re}X_4 \right) - e^A \text{Re}X_1 \wedge G_4 = 0. \label{eq:BPSAdS4Neq1i}
\end{align}
\end{subequations}
In these conditions we assumed $m \neq 0$, in appendix \ref{app:Neq1} we provide the full set of conditions that include the $m=0$, \textit{i.e.} Mink$_{4}$, limit.

\subsection{Parametrising bilinears in terms of SU(3)-structure forms}

We first decompose $\chi$ in terms of two Majorana spinors $\chi_{1,2}$ as
\begin{equation}
\chi = e^{\frac{A}{2}} (\chi_1 + i \chi_2).
\end{equation}
Following appendix \ref{app:canonical frame}, we then further decompose $\chi_{1,2}$ in terms of a single unit norm Majorana spinor $\eta$ and a vielbein direction $V$ as
\begin{equation}\label{eq:M7decspinor}
\chi_1 = a_1 \eta, \qquad \chi_2 = a_2 \eta - i a_3 V \eta,
\end{equation}
where $a_I$ are functions obeying $a_1^2 + a_2^2 + a_3^2 = 1$ such that the norm of $\chi$ is $e^A$ \eqref{eq:defXNeq1}. In terms of this parametrisation the bilinears decompose as
\begin{equation}\label{eq:bilinearsparamNeq1}
\begin{split}
Y_0 & = (a_1^2 - a_2^2 - a_3^2) e^A + 2 i a_1 a_2 e^A, \qquad X_1 = 2 e^A a_1 a_3 V, \qquad X_2 = 2 i a_1 a_3 e^A J, \\[2mm]
Y_3 & = - 2 e^A a_1 (a_2 J \wedge V + a_3 \text{Re}\Omega - a_2 \text{Im}\Omega) \\[2mm]
& + i e^A ( (a_1^2 - a_2^2 - a_3^2) J \wedge V - 2 a_2 a_3 \text{Re}\Omega - (a_1^2 -a_2^2 + a_3^2) \text{Im}\Omega ), \\[2mm]
X_4 & = - \frac{e^A}{2} \left( J \wedge J + 2 (a_1^2 + a_2^2 - a_3^2) \text{Re}\Omega \wedge V + 4 a_2 a_3 \text{Im}\Omega \wedge V \right),
\end{split}
\end{equation}
where $J$ and $\Omega$ are orthogonal to $V$ and span an SU($3$)-structure, \textit{i.e.}
\begin{equation}
J \wedge \Omega = 0, \qquad J \wedge J \wedge J = \frac{3}{4} i \Omega \wedge \bar{\Omega}.
\end{equation}

Before moving to the next section, it is useful to comment on several special tunings of the parameters $a_i$. The details of these cases are discussed in appendix \ref{app:Neq1}:
\begin{itemize}
\item[-] $a_1 \neq 0$. Otherwise one can show the corresponding class dies.
\item[-] $a_ 2 = 0 \Rightarrow c=0$. This corresponds to the class of solutions describing M$5$ branes wrapping cycles, see \cite{Gauntlett:2006ux,Pernici:1984nw,Acharya:2000mu,Gauntlett:2000ng} for example.
\item[-] $a_3 = 0$, consistency then requires $ (a_1,a_2) \in \{ \pm \frac{1}{\sqrt{2}} \}$, this leads to the well known class of AdS$_4$ solutions of Freund-Rubin type for which M$_7$ is constrained to be a weak G$_2$-manifold.
\end{itemize}

\subsection{Integrability}

From the analysis in \cite{Prins:2013wza}, we can state that the supersymmetry conditions \eqref{eq:BPSAdS4Neq1} combined with the flux equations and Bianchi identity \eqref{eq:fluxBianchieqs7d} imply all remaining equations of motion. In fact, we can make a stronger statement on the flux equations \eqref{eq:fluxBianchieqs7d} being implied by the supersymmetry conditions. We observe that \eqref{eq:fluxeq7d} is provided by a suitable linear combination of \eqref{eq:BPSAdS4Neq1}, namely
\begin{equation}
3 m \eqref{eq:BPSAdS4Neq1g} - d \eqref{eq:BPSAdS4Neq1f} - \eqref{eq:BPSAdS4Neq1a} G_4 = \eqref{eq:fluxeq7d}.
\end{equation}
We conclude that \eqref{eq:fluxeq7d} is implied by supersymmetry. One can derive constraints on \eqref{eq:Bianchieq7d} as follows
\begin{align}
& d \eqref{eq:BPSAdS4Neq1g} - \frac{1}{3 m} \eqref{eq:BPSAdS4Neq1a} \wedge G_4 = e^{2A} \text{Im} Y_0 d G_4 , \label{eq:Bianchi1} \\[2mm]
& 2 m \eqref{eq:BPSAdS4Neq1i} + d \eqref{eq:BPSAdS4Neq1h} - d \eqref{eq:BPSAdS4Neq1b} \wedge G_4 = e^{2A} \text{Re} Y_0 d G_4. \label{eq:Bianchi2}
\end{align}
Clearly supersymmetry then implies \eqref{eq:Bianchieq7d} locally, unless $\text{Re}Y_0 = 0$ or $\text{Im}Y_0 = 0$ globally - however, as we discuss at the end of appendix \ref{app:Neq1}, this is not possible. Sources are still possible, but these must happen at local points in the space where $e^{2A}Y_0 dG_4 = 0$.

\section{${\cal{N}} > 1 $ AdS$_4$ conditions} \label{sec:Neq123}

In this section we provide the necessary and sufficient conditions for AdS$_4$ solutions of eleven dimensional supergravity to preserve ${\cal{N}}=3$ supersymmetry. These conditions contain analogous conditions to realise both ${\cal{N}}=2$ and ${\cal{N}}=1$ supersymmetry as special cases. This section is a summary of appendix \ref{app:Neq3}, we refer the reader to it for more details.
~\\

In order to preserve ${\cal{N}}=3$ supersymmetry, we need a solution of the form \eqref{eq:Ansatz11dbosonic} that supports three independent ${\cal{N}}=1$ spinors of the form \eqref{eq:spinor11DNeq1}. In other words \eqref{eq:spinor11DNeq1} is generalised to
\begin{equation}\label{eq:spinor11D}
\epsilon = \zeta^I_+ \otimes \chi^I + \zeta^I_- \otimes \chi^{I c},
\end{equation}
for $I= \{1, 2, 3 \}$, which is again Majorana by construction. Similarly to the previous section, $\chi^I$ are three independent Dirac spinors with support on M$_7$ that must be non-zero to realise ${\cal{N}} = 3$, while $\zeta^I$ are now three Killing spinors on AdS$_4$ that satisfy the Killing spinor equation
\begin{equation}\label{eq:KSEAdS4chiral}
\nabla_{\mu}^{(4)} \zeta^I_{\pm} = \frac{m}{2} \gamma_{\mu}^{(4)} \zeta^I_{\mp},
\end{equation}
which is the analogue of \eqref{eq:KSEAdS4chiralNeq1}. We refer to appendix \ref{app:Neq3} for further details and for the explicit parametrisation of the spinors.
~\\

Similar to the analysis in the previous section, one can construct bilinear conditions on the internal manifold M$_7$ that imply ${\cal{N}}=3$ supersymmetry starting from \eqref{eq:spinor11D} and \eqref{eq:KSEAdS4chiral}. We derive these in appendix \ref{app:Neq3}, and present a summary of that derivation here.

We first define the bilinears $X_n^{IJ}$ and $Y_n^{IJ}$ constructed from $\chi^I$, in terms of which the supersymmetry conditions are expressed
\begin{equation}\label{eq:XIJYIJ}
X^{IJ}_{n} = \frac{1}{n !} \chi^{I \dagger} \gamma_{a_1 ... a_n} \chi^{J} e^{a_1 ... a_n}, \qquad Y^{IJ}_{n} = \frac{1}{n !} \chi^{I c \dagger} \gamma_{a_1 ... a_n} \chi^{J} e^{a_1 ... a_n},
\end{equation}
where the symmetries properties of these bilinears can be found in \eqref{eq:symmXIJYIJ} and \eqref{eq:antisymmXIJYIJ}. The first outcome of this derivation is that in general, AdS$_4$ solutions experience an enhancement to AdS$_5$ unless one imposes
\begin{subequations}\label{eq:noAdS5}
\begin{align}
& \text{Re} X_0^{IJ} = e^A \delta^{IJ}, \label{eq:noAdS5a} \\[2mm]
& \text{Im} Y^{IJ}_1 = \text{Im} X_0^{IJ} = 0. \label{eq:noAdS5b}
\end{align}
\end{subequations}
Since we are interested in AdS$_4$ solutions, \eqref{eq:noAdS5} must be imposed. Furthermore, as a second outcome, one derives the following
\begin{align}
&\nabla^{(7)}_{(a} \text{Re} (Y^{IJ}_1)_{b)}=0, \label{eq:7dsusyantisymm0} \\[2mm] 
&\mathcal{L}_{\text{Re} Y^{IJ}_1 } A = 0, \label{eq:7dsusyantisymm0s}
\end{align}
which implies that $\text{Re} (Y^{IJ}_1)$ is spanned by three independent Killing vectors of the eleven dimensional manifold\footnote{We remark that Re$Y^{IJ}_1$ is antisymmetric in $I,J$, so it has 3 independent components, see \eqref{eq:antisymmXIJYIJ}.}, this is consistent with the SO(3) R-symmetry that the internal space should support when ${\cal{N}}=3$ supersymmetry is preserved.

As shown in appendix \ref{app:Neq3}, the internal spinors transform non-trivially under the action of these Killing vectors, \textit{i.e.} the Lie derivative of these spinors along the Killing vectors gives
\begin{equation}
{\cal{L}}_{\text{Re} Y_1^{IJ}} \chi^K = m ({\cal{S}}^{IJ})^{K L} \chi^L,
\end{equation}
where ${\cal{S}}^{IJ}$ are the SO($3$) generators of the R-symmetry Lie algebra \eqref{eq:defcalS} -- as such $\chi^I$ transform as a triplet under the SO($3$) R-symmetry. One has that $X^{IJ}_n, Y^{IJ}_n$ decompose as the product $ \mathbf{3} \otimes \bar{\mathbf{3}} = \mathbf{1} \oplus \mathbf{3} \oplus \mathbf{5}$, under SO($3$). Here, the $\mathbf{1}$ corresponds to the trace, the $\mathbf{3}$ to the antisymmetric, and the $\mathbf{5}$ to the symmetric and traceless components of the $3 \times 3$ matrices $X^{IJ}_n, Y^{IJ}_n$. Given this, one can derive the necessary and sufficient conditions for supersymmetric ${\cal{N}} = 3$ AdS$_4$. These split into two groups, depending on whether they are symmetric or antisymmetric in the indices $I,J$. The symmetric conditions are
\begin{subequations}\label{eq:7dsusysymmsec2}
\begin{align}
& d (e^{2 A} \text{Re} Y_0^{IJ}) - 2 m e^A \text{Re} X_1^{IJ} = 0, \label{eq:7dsusysymma} \\[2mm]
& 3 m e^{2 A} \text{Im}  Y_0^{IJ} - c \delta^{IJ} = 0, \label{eq:7dsusysymmb} \\[2mm]
& 6 e^A \delta^{IJ} \star_7 dA - \text{Im} X_2^{IJ} \wedge G_4 = 0, \label{eq:7dsusysymmc} \\[2mm]
& (6 e^{4A} m \delta^{IJ} - 2 c \text{Im} Y_0^{IJ} ) \text{vol}_7 - e^{4A} \text{Im} Y_3^{IJ} \wedge G_4 = 0, \label{eq:7dsusysymmd} \\[2mm]
& 2 c \text{Re} Y_0^{IJ} \text{vol}_7 + e^{4A} \text{Re} Y_3^{IJ} \wedge G_4 = 0, \label{eq:7dsusysymme} \\[2mm]
& d (e^{3A} \text{Im} X_2^{IJ} )+ 3 m e^{2 A} \text{Im} Y_3^{IJ}- e^{4A} \delta^{IJ} \star_7 G_4 = 0, \label{eq:7dsusysymmf} \\[2mm]
& d (e^{2A} \text{Re} Y_3^{IJ} ) - 2 m e^A \text{Re} X_4^{IJ} + e^{2A} \text{Re} Y_0^{IJ} G_4 = 0, \label{eq:7dsusysymmg} \\[2mm]
& d (e^{2A} \text{Im} Y_3^{IJ} ) + e^{2A} \text{Im} Y_0^{IJ} G_4 = 0, \label{eq:7dsusysymmh} \\[2mm]
& d (e^A \text{Re} X_4^{IJ} ) - e^A \text{Re} X_1^{IJ} \wedge G_4 = 0, \label{eq:7dsusysymmi}
\end{align}
\end{subequations}
while the antisymmetric ones are
\begin{subequations}\label{eq:7dsusyantisymmsec2}
\begin{align}
& d (e^A \text{Im} X_1^{IJ}) + m \text{Im}Y_2^{IJ} = 0, \label{eq:7dsusyantisymma} \\[2mm]
& d (\text{Re} Y_2^{IJ}) - i_{\text{Re}Y_1^{IJ}} G_4 = 0, \label{eq:7dsusyantisymmb} \\[2mm]
& 3 e^{4A} \star_7 d \text{Re} Y_1^{IJ} + c \text{Re} Y_5^{IJ} - 2 e^{4A} \text{Re} Y_2^{IJ} \wedge \star_7 G_4 = 0, \label{eq:7dsusyantisymmc} \\[2mm]
& 2 e^{4A} \text{Im} Y_2^{IJ} \wedge \star_7 G_4 - e^{4A} \text{Re} Y_1^{IJ} \wedge G_4 - c \text{Im} Y_5^{IJ} = 0, \label{eq:7dsusyantisymmd} \\[2mm]
& \text{Re}X_2^{IJ} \wedge G_4 = 0, \label{eq:7dsusyantisymme} \\[2mm]
& d \left( e^{4A} \text{Re} Y_1^{IJ} \right) - 4 m e^{3A} \text{Re}X_2^{IJ} + c \text{Im} Y_2^{IJ} = 0, \label{eq:7dsusyantisymmf} \\[2mm]
& e^{4A} i_{\text{Re} Y_1^{IJ}} \star_7 G_4 - c \text{Re} Y_2^{IJ} = 0, \label{eq:7dsusyantisymmg} \\[2mm]
& d (e^A \text{Im} X_4^{IJ}) - e^A \text{Im} X_1^{IJ} \wedge G_4 + m \text{Im} Y_5^{IJ} = 0, \label{eq:7dsusyantisymmh} \\[2mm]
& d(\text{Re} Y_5^{IJ}) + c e^{- 4 A} i_{\text{Re} Y_1^{IJ} } \text{vol}_7 + \text{Re} Y_2^{IJ} \wedge G_4 = 0, \label{eq:7dsusyantisymmi} \\[2mm]
& d(\text{Im} X_5^{IJ}) + \text{Im} Y_2^{IJ}\wedge G_4 = 0. \label{eq:7dsusyantisymmj}
\end{align}
\end{subequations}
Note that these conditions assume that $m \neq 0$ and that the no-AdS$_5$ of \eqref{eq:noAdS5}, which are only necessary in this case, have been imposed. In appendix \ref{app:Neq3} we provide the full conditions that admit the limit $m=0$.

The diagonal components of \eqref{eq:7dsusysymmsec2} give three copies of the ${\cal{N}} = 1$ system of the previous section which already provide necessary and sufficient conditions for supersymmetry. The rest of \eqref{eq:7dsusysymmsec2} and \eqref{eq:7dsusyantisymmsec2} are redundant, as far as sufficiency is concerned, but they are very useful for actually solving the system.

Note that one can combine \eqref{eq:7dsusyantisymmb} and \eqref{eq:Bianchieq7d} to establish that
\begin{equation}
{\cal{L}}_{\text{Re}Y^{IJ}_1} G_4=0,
\end{equation}
thus the SO(3) isometry of the metric is also respected by the fluxes.

\section{Recovering the ${\cal{N}}=2$ class}\label{sec:Neq2class}

Before moving onto the classification of ${\cal N}=3$ solutions, we will perform a non-trivial check of the supersymmetry constraints we derived in the previous section. Specifically we will recover a class of ${\cal{N}}=2$ solutions with both electric and magnetic four-form flux components first derived in \cite{Gabella:2012rc}. Breaking supersymmetry from ${\cal{N}}=3$ to ${\cal{N}}=2$ in \eqref{eq:spinor11D} just amounts to setting one component of $\chi^I$ to zero, which, in terms of \eqref{eq:7dsusysymmsec2} and \eqref{eq:7dsusyantisymmsec2}, is equivalent to restricting to $ I \in \{1,2\}$ - there is then only a single Killing vector contained in Re$Y_1^{IJ}$, consistent with the SO(2) R-symmetry one expects for ${\cal{N}}=2$.~\\

We now set about recovering the class from \cite{Gabella:2012rc}: To this end, we decompose the spinor in the following fashion
\begin{equation}\label{eq:spinordecompos}
\chi^I = \frac{1}{\sqrt{2}}(\eta^I_+ + \eta^I_-) + i \frac{1}{\sqrt{2}}(\eta^I_+ - \eta^I_-),
\end{equation}
where $\eta^I_{\pm}$ are Majorana. To solve the system we first focus on the zero- and one-form constraints \eqref{eq:noAdS5}, \eqref{eq:7dsusysymmb}. One can show that these can be solved without loss of generality by taking
\begin{equation}
\eta_{\pm}^1 = \lambda_{\pm} \eta_{\pm}, \qquad \eta_{\pm}^2 = - i \lambda_{\pm} W_{\pm} \eta_{\pm},
\end{equation}
where $\lambda_{\pm}$ are two functions, $W_{\pm}$ are unit norm, and $\eta_{\pm}$ are two unit norm Majorana spinors defined on M$_7$. We can then expand $\eta_{\pm}$ in a common basis as
\begin{equation}
\eta_+ = \eta, \qquad \eta_- = \cos \alpha \eta - i \sin \alpha V \eta,
\end{equation}
for $\eta$ a unit norm Majorana spinor, that can be taken to be in the canonical frame described in appendix \ref{app:canonical frame}. Here $\alpha$ is a function of M$_7$ and $V$ a unit norm one-form on M$_7$. Solving \eqref{eq:noAdS5} and \eqref{eq:7dsusysymmb} requires the one-forms $(V, W_{\pm})$ to be expressed in terms of a vielbein $e^a$ on M$_7$ as
\begin{equation}
\begin{aligned}
V & = e^7, \qquad W_+ = \cos \beta e^7 + \sin \beta e^6, \\[2mm]
W_- & = - \cos \beta e^7 + (\sin^2 \alpha - \cos^2 \alpha) \sin \beta e^6 + 2 \cos \alpha \sin \alpha \sin \beta e^5,
\end{aligned}
\end{equation}
where $\beta$ is a function on M$_7$ and
\begin{equation}
\lambda_{\pm} = \sqrt{\frac{e^A}{2} \pm \frac{c}{6 e^{2A} m }}.
\end{equation}
We first note that $\cos \alpha, \sin \alpha$ and $\cos \beta$ cannot be zero, otherwise \eqref{eq:7dsusysymma} cannot be solved, unless one focuses on the Sasaki-Einstein case where $\lambda_+ \lambda_- = 0$. Furthermore, $\sin \beta$ is not allowed to be zero, otherwise one can show that \eqref{eq:7dsusysymma}-\eqref{eq:7dsusyantisymma} imply $m=0$. 

We want to introduce a different parametrisation of the two angles $\alpha, \beta$ to make the connection with \cite{Gabella:2012rc} easier. To this end we define two arbitrary functions on M$_7$ $(\rho, \chi)$ by identifying  
\begin{equation}
e^{2A} m (\text{Re}X_0^{12} + i \text{Re}X_0^{11})= \frac{1}{3} \sqrt{9 e^{6A} m^2 - c^2} (i \cos \alpha + \cos \beta \sin \alpha) = \rho e^{i \chi},
\end{equation}
and identify three more useful vielbein directions $E_i$ by means of the one-form bilinears,
\begin{align}\label{eq:defvielbeine}
\text{Re}X_1^{11} & = - \text{Re}X_1^{22} = \frac{\rho \cos \chi}{e^{2 A} m \sqrt{\Delta}} E_1 - e^{A} \sqrt{1- \Delta} \sin \chi E_2 - \frac{c \sqrt{1 - \Delta} \cos \chi}{3 e^{2 A} m \sqrt{\Delta}} E_3, \nn \\[2mm]
\text{Re}X_1^{12} & = - \frac{ \rho \sin \chi}{e^{2 A} m \sqrt{\Delta}} E_1 - e^{A} \sqrt{1- \Delta} \cos \chi E_2 + \frac{c \sqrt{1 - \Delta } \sin \chi}{3 e^{2 A} m \sqrt{\Delta}} E_3, \\[2mm]
\text{Re}Y_1^{12} & = - e^A \sqrt{\Delta} E_1. \nn
\end{align}
The parametrisation of $E_i$ in terms of $e^5, e^6, e^7$ is given in appendix \ref{sec:AppendixNeq2}, while $\Delta$ is given by
\begin{equation}
\Delta = \frac{c^2 + 9 \rho^2}{9 e^{6A} m^2}. 
\end{equation}
As explained in more detail in appendix \ref{subapp:SU3SU2}, $E_i$ identify an SU(2)-structure within the internal space as
\begin{equation}
ds^2(\text{M}_7) = ds^2(\text{SU}(2)) + E_i^2,
\end{equation}
with three real SU(2)-structure two-forms $j_a$ which obey $ j_a \wedge j_b = 2 \delta_{ab} \text{vol}_4$ and define $ds^2(\text{SU}(2))$.
To determine the explicit form of the vielbein $E_i$, we first observe that a linear combination of \eqref{eq:7dsusysymma} provides a direct constraint that fixes $E_2$, namely
\begin{equation}
0 = \frac{ \sin \chi \eqref{eq:7dsusysymma}^{11} + \cos \chi \eqref{eq:7dsusysymma}^{12} }{3}  = d \rho + 2 e^{2A} m^2 \sqrt{1-\Delta} E_2,
\end{equation}
where we employ notation such that \eqref{eq:7dsusysymma}$^{12}$ indicates the part of \eqref{eq:7dsusysymma} with $(I,J)=(1,2)$ and so on. We thus see that we can take $\rho$ to define a coordinate on M$_7$. From \eqref{eq:defvielbeine} and \eqref{eq:11dKilling} it is clear that $E_1$ is parallel to the Killing vector in Re$Y^{IJ}_1$, therefore we can introduce local coordinates such that $E_1$ takes the form
\begin{equation}
E_1 = - \frac{e^A \Delta}{2 m} ( d \psi + {\cal{A}}),
\end{equation}
where $\partial_{\psi}$ spans the SO(2) Killing vector. Thus $e^{A}$, $\Delta$ and ${\cal{A}}$ are independent of $\psi$. To determine the last vielbein direction $E_3$, we make use of the combination
\begin{equation}
0 = \frac{\cos \chi \eqref{eq:7dsusysymma}^{11} - \sin \chi \eqref{eq:7dsusysymma}^{12}}{3 \rho}  = d \chi - \frac{2}{e^A \sqrt{\Delta}} \left( m E_1 - \frac{c}{9 e^{3A} \rho} \sqrt{ 9 e^{6A} m^2 -c^2 - 9 \rho^2} E_3 \right),
\end{equation}
which implies
\begin{equation}
E_3 = - \frac{3 e^A \rho \sqrt{\Delta}}{2 c m \sqrt{1 - \Delta}} (d \tau + {\cal{A}}),
\end{equation}
upon fixing $\chi = \psi - \tau$. We can thus take $\tau$ to define another coordinate, arriving at the three coordinates spanning $E_i$. At this stage, the only thing left to fully fix $E_i$ is to determine the field strength of the connection ${\cal{A}}$. In this regard, the $(1,2)$ component of \eqref{eq:7dsusyantisymma} can be written as
\begin{equation}\label{eq:Neq2eq1}
d \left( \frac{1}{3 m e^A \Delta} \left(c E_1 + 3 \rho \sqrt{1- \Delta} E_3 \right) \right) + e^A m \left( j_3 - \sqrt{\Delta} E_2 \wedge E_3\right) = 0.
\end{equation}
This provides the following definition of the field strength
\begin{equation}
d {\cal{A}} = \frac{2 c m}{3 e^{5A} \Delta} \left( j_3 + \left(3 \sqrt{\Delta} - \frac{4}{\sqrt{\Delta}} \right) E_2 \wedge E_3 \right).
\end{equation}
To fully recover the class of \cite{Gabella:2012rc}, we still need to derive two of the differential constraints and the definition of $G_4$. In \cite{Gabella:2012rc} the authors were able to derive an expression for $\star_7 G_4$, then taking the hodge dual of it provides $G_4$. However, thanks to our formalism, we are able to directly derive a condition on $G_4$. One can show that \eqref{eq:7dsusysymmg}$^{11}$,\eqref{eq:7dsusysymmg}$^{12}$,\eqref{eq:7dsusysymmg}$^{22}$,\eqref{eq:7dsusysymmh}$^{11}$,\eqref{eq:7dsusysymmh}$^{22}$ and the one-form constraints introduced earlier in this section can be reduced to the following equations
\begin{align}
& d \left( e^{6A} \Delta E_2 \wedge j_2 \right) - \rho \, d\left( e^{6A} \rho^{-1} \sqrt{\Delta} j_1 \wedge E_3\right) = 0, \\[2mm]
& d \left( e^{3A}  E_2 \wedge j_1 \right) + \rho \, d \left( e^{3A} \rho^{-1} \sqrt{\Delta}  j_2 \wedge E_3 \right) = 0, \\[2mm]
& G_4 = \frac{1}{e^A \sqrt{\Delta} } E_1 \wedge d \left( e^A \sqrt{1 - \Delta}j_1 \right) + \frac{c \sqrt{1 - \Delta}}{ e^{4A} \sqrt{\Delta} } E_2 \wedge E_3 \wedge j_1.
\end{align}
These, along with the local forms of $E_i$ and \eqref{eq:Neq2eq1}, are the necessary and sufficient conditions, for a supersymmetric solution. One can check that all the other constraints in \eqref{eq:7dsusysymmsec2} and \eqref{eq:7dsusyantisymmsec2} are implied by these.

We find complete agreement with the expressions derived in \cite{Gabella:2012rc} provided that
\begin{equation}
A = \hat{\Delta}, \qquad E_i = e^{\hat{\Delta}} \hat{E}_i, \qquad m = 2, \qquad c = -\hat{m}, \qquad \text{vol}_4 = -\hat{\text{vol}}_4, \qquad \Delta = \hat{||\xi||}^2,
\end{equation}
where hats indicate objects in \cite{Gabella:2012rc} and $j_i$ and ${\cal{A}}$ are the same here and there.
~\\

In summary, we recover the class from \cite{Gabella:2012rc}, which in our conventions decomposes as
\begin{align}
& ds^2 = e^{2A} ds^2(\text{AdS}_4) + ds^2(\text{M}_7), \qquad ds^2(\text{M}_7) = ds^2(\text{SU}(2)) + E_i^2, \\[2mm]
&  G = c \, \text{vol}(\text{AdS}_4) + G_4 \qquad G_4 = \frac{1}{e^A \sqrt{\Delta} } E_1 \wedge d \left( e^A \sqrt{1 - \Delta}j_1 \right) + \frac{c \sqrt{1 - \Delta}}{ e^{4A} \sqrt{\Delta} } E_2 \wedge E_3 \wedge j_1. \nn
\end{align}
Supersymmetry demands that the following conditions are obeyed
\begin{equation}
\begin{aligned}
& d \left( \frac{1}{3 m e^A \Delta} \left(c E_1 + 3 \rho \sqrt{1- \Delta} E_3 \right) \right) + e^A m \left( j_3 - \sqrt{\Delta} E_2 \wedge E_3\right) = 0, \\[2mm]
& d \left( e^{6A} \Delta E_2 \wedge j_2 \right) - \rho \, d\left( e^{6A} \rho^{-1} \sqrt{\Delta} j_1 \wedge E_3\right) = 0, \\[2mm]
& d \left( e^{3A}  E_2 \wedge j_1 \right) + \rho \, d \left( e^{3A} \rho^{-1} \sqrt{\Delta} j_2 \wedge E_3 \right) = 0.
\end{aligned}
\end{equation}
One has a supersymmetric solution whenever these conditions are satisfied.

\section{Deriving the ${\cal{N}}=3$ class}\label{sec:Neq3class}

In this section, we classify ${\cal{N}}=3$ supersymmetric AdS$_4$ solutions in eleven dimensional supergravity. A well known family of such solutions is given by the Tri-Sasaki backgrounds \cite{Acharya:1998db}, see \cite{Boyer:1998sf} for a review of this geometry. As the name suggests, the internal manifolds of these solutions are Tri-Sasaki, and the resulting backgrounds are of Freund-Rubin type. A Tri-Sasaki manifold is an S$^3$ bundle fibred over a quaternionic-K\"ahler base, with the total space being Einstein. We provide more details in the following section. This class has been known for a long time and has provided many results and applications \cite{Acharya:1998db,PandoZayas:2015auw,Cassani:2011fu,Jafferis:2008qz,Castellani:1983tc,Termonia:1999cs,Fre:1999gok,Billo:2000zs}. We begin by solving the zero-form constraints \eqref{eq:noAdS5} and \eqref{eq:7dsusysymmb}, which lead to a branching into two cases. In section \ref{subsec:TriSasaki}, we show that the first branch recovers the Tri-Sasaki class, while in section \ref{sec:nootherNeq3} we prove that the second branch is ruled out. This establishes that the Tri-Sasaki structure is the unique class preserving ${\cal{N}}=3$ supersymmetry.
~\\

Similar to the previous section, we begin by parametrising the spinor as
\begin{equation}\label{eq:chiINeq3}
\chi^I = \frac{1}{\sqrt{2}}(\eta^I_+ + \eta^I_-) + i \frac{1}{\sqrt{2}}(\eta^I_+ - \eta^I_-),
\end{equation}
where $\eta^I_{\pm}$ are Majorana and now $I \in \{1,2,3\}$. From \eqref{eq:noAdS5} and \eqref{eq:7dsusysymmb} we find the following conditions
\begin{subequations}\label{eq:0and1formeqs}
\begin{align}
& \eta^{1 \dagger}_{\pm} \eta^{1}_{\pm} = \eta^{2 \dagger}_{\pm} \eta^{2}_{\pm} = \eta^{3 \dagger}_{\pm} \eta^{3}_{\pm} = \lambda_{\pm}^2, \label{eq:0and1formeqsa} \\[2mm]
& \eta^{1 \dagger}_{\pm} \eta^{2}_{\pm} = \eta^{1 \dagger}_{\pm} \eta^{3}_{\pm} = \eta^{2 \dagger}_{\pm} \eta^{3}_{\pm} = 0, \label{eq:0and1formeqsb} \\[2mm]
& \eta^{2 \dagger}_+ \eta^1_- - \eta^{2 \dagger}_- \eta^{1 \dagger}_+ = \eta^{3 \dagger}_+ \eta^1_- - \eta^{3 \dagger}_- \eta^{1 \dagger}_+ = \eta^{3 \dagger}_+ \eta^2_- - \eta^{3 \dagger}_- \eta^{2}_+ = 0, \label{eq:0and1formeqsc} \\[2mm]
& \text{Im} (\eta^{1 \dagger}_{+} \gamma_a \eta^{2}_{-} + \eta^{1 \dagger}_- \gamma_a \eta^{2}_+ ) e^a = 0, \label{eq:0and1formeqsd} \\[2mm]
& \text{Im} (\eta^{1 \dagger}_{+} \gamma_a \eta^{3}_{-} + \eta^{1 \dagger}_- \gamma_a \eta^{3}_+ ) e^a = 0, \label{eq:0and1formeqse} \\[2mm]
& \text{Im} (\eta^{2 \dagger}_{+} \gamma_a \eta^{3}_{-} + \eta^{2 \dagger}_- \gamma_a \eta^{3}_+ ) e^a = 0, \label{eq:0and1formeqsf}
\end{align}
\end{subequations}
where \begin{equation}\label{eq:deflambdapm}
\lambda_{\pm}^2 = \frac{e^A}{2} \pm \frac{c}{6 e^{2A} m }.
\end{equation}
At this point, we must distinguish between the case in which either $\lambda_+ = 0$ (or, analogously, $\lambda_- = 0$) and the case in which neither vanishes. These define two distinct branches, which we analyse separately. The former corresponds to the Tri-Sasaki limit, so we begin by deriving this case.

\subsection{Tri-Sasaki class}\label{subsec:TriSasaki}

In this subsection we focus on recovering the Tri-Sasaki class. In particular, this is achieved by setting one of $\lambda_{\pm}$ in \eqref{eq:deflambdapm} to zero, which can be accomplished by choosing $c = \pm 3 m e^{3A} $. We can choose without loss of generality which of the two we set to zero. In what follows, we focus on 
\begin{equation}\label{eq:lambdacforG2}
\lambda_+^2 = e^A, \qquad \lambda_- = 0, \qquad  c = 3 m e^{3A},
\end{equation}
this is equivalent to setting $\eta^I_- = 0$. Since, for this class, the warp factor $e^A$ is constant, we implement the following redefinition $e^A = L m$, that allows us to write the fields as
\begin{equation}
ds^2 = L^2 \left( m^2 ds^2(\text{AdS}_4) + 4 ds^2( \hat{ \text{M}}_7) \right), \qquad ds^2(\text{M}_7) = 4 L^2 ds^2(\hat{{\text{M}}}_7),
\end{equation}
making AdS$_4$ effectively of unit radius. Using \eqref{eq:lambdacforG2}, the spinors \eqref{eq:chiINeq3} become
\begin{equation}
\chi^I = \frac{1+i}{\sqrt{2}} \eta^I,
\end{equation}
where $\eta^I = \eta^I_+$, from now on we drop the superscript $+$ since it is the only spinor that appears. In this limit, \eqref{eq:0and1formeqs} simplify to
\begin{subequations}\label{eq:0andformeqG2case}
\begin{align}
& \eta_1^{\dagger} \eta_1 = \eta_2^{\dagger} \eta_2 = \eta_3^{\dagger} \eta_3 = e^A, \\[2mm]
& \eta_1^{\dagger} \eta_2 = \eta_1^{\dagger} \eta_3 = \eta_2^{\dagger} \eta_3 = 0.
\end{align}
\end{subequations}
These constrains are solved without loss of generality by taking the following parametrisation of the three spinors
\begin{equation}
\eta^1 = \sqrt{L m} \eta, \qquad \eta^2 = - i \sqrt{L m}  W \eta, \qquad \eta^3 = - i \sqrt{L m}  Z \eta,
\end{equation}
where $W$ and $Z$ are two one-forms which must obey
\begin{equation}\label{eq:WZG2}
i_{W} W = i_Z Z = 1, \qquad i_W Z = 0.
\end{equation}
We choose without loss of generality to align these along two vielbein directions as
\begin{equation}
W = e^7, \qquad Z = e^6,
\end{equation}
where $e^a$ span a vielbein on M$_7$ as before. We now define the following vielbein basis on $\hat{\text{M}}_7$ and real SU(2)-structure two-forms
\begin{equation}
\begin{aligned}
\text{Re}Y^{IJ}_1 & = 2 m L^2
\begin{pmatrix}
0 & \hat{E}_3 & \hat{E}_2 \\
- \hat{E}_3 & 0 & \hat{E}_1 \\
- \hat{E}_2 & \hat{E}_1 & 0
\end{pmatrix}, \\[2mm]
\text{Re}X^{IJ}_2 & = 4 L^3 m
\begin{pmatrix}
0 & \hat{j}_3 + \hat{E}_1 \wedge \hat{E}_2 & \hat{j}_2 - \hat{E}_1 \wedge \hat{E}_3 \\
- \hat{j}_3 - \hat{E}_1 \wedge \hat{E}_2 & 0 & -\hat{j}_1 - \hat{E}_2 \wedge \hat{E}_3 \\
- \hat{j}_2 + \hat{E}_1 \wedge \hat{E}_3 & \hat{j}_1 + \hat{E}_2 \wedge \hat{E}_3 & 0
\end{pmatrix}.
\end{aligned}
\end{equation}
where the $\hat{E}_i$ span directions orthogonal to the $\hat{j}_a$. In terms of these, we can write $ds^2(\hat{\text{M}}_7)$ as
\begin{equation}
ds^2(\hat{\text{M}}_7) = ds^2(\text{M}_{QK}) + \hat{E}_i^2.
\end{equation}
where the $\hat{j}_i$ have legs entirely along M$_{QK}$ and they obey $\hat{j}_i \wedge \hat{j}_j = 2 \delta_{ij} \text{vol}(\text{M}_{QK})$. With respect to the explicit vielbein, we have
\begin{equation}
\hat{E}_i = \frac{1}{2 L} e^{i+4}, \qquad \hat{j}_i = \frac{1}{4 L^2} \tilde{j}_i,
\end{equation}
where $\tilde{j}_i$ have been defined in appendix \ref{app:canonical frame}. Upon some manipulation, one can show that \eqref{eq:7dsusyantisymma} is solved by
\begin{equation}\label{eq:Eitri}
d \hat{E}_i = 2 \hat{j}_i + \frac{1}{2} \epsilon_{ijk} \hat{E}_j \wedge \hat{E}_k.
\end{equation}
Solving \eqref{eq:7dsusysymmi} also requires 
\begin{equation}\label{eq:jitri}
d \hat{j}_i = - 2 \epsilon_{ijk} \hat{j}_j \wedge \hat{E}_k.
\end{equation}
One can check that all of \eqref{eq:7dsusysymmsec2} and \eqref{eq:7dsusyantisymmsec2} are implied by \eqref{eq:Eitri} and \eqref{eq:jitri}.
~\\

In summary, we recover the well known class of solutions with Tri-Sasaki internal spaces, in our conventions these take the form
\begin{equation}
ds^2 = L^2 \left( m^2 ds^2(\text{AdS}_4) + 4 (ds^2(\text{M}_{QK}) + \hat{E}_i^2) \right), \qquad G = c \, \text{vol}(\text{AdS}_4),
\end{equation}
where a solution must obey
\begin{equation}
d \hat{E}_i = 2 \hat{j}_i + \frac{1}{2} \epsilon_{ijk} \hat{E}_j \wedge \hat{E}_k, \qquad d \hat{j}_i = - 2 \epsilon_{ijk} \hat{j}_j \wedge \hat{E}_k,
\end{equation}
for $\hat{j}_i$ the real SU(2)-structure forms on the quaternionic-K\"ahler manifold M$_{QK}$.

\subsection{${\cal{N}}=3$ implies Tri-Sasaki} \label{sec:nootherNeq3}

In this section we establish that ${\cal{N}} > 2$ implies Tri-Sasaki internal space for AdS$_4$.

~\\
To solve \eqref{eq:0and1formeqs} in full generality when $\lambda_{\pm} \neq 0$, $\eta^I_{\pm}$ must decompose in terms of two unit norm Majorana spinors $\eta_{\pm}$ as
\begin{equation}
\eta^1_{\pm} = \lambda_{\pm} \eta_{\pm}, \qquad \eta^2_{\pm} = - i \lambda_{\pm} W_{\pm} \eta_{\pm}, \qquad \eta^3_{\pm} = - i \lambda_{\pm} Z_{\pm} \eta_{\pm},
\end{equation}
where $W_{\pm}$ and $Z_{\pm}$ are one-forms such that
\begin{equation}\label{eq:eqsWpmZpm0}
i_{W_{\pm}} W_{\pm} = i_{Z_{\pm}} Z_{\pm} = 1, \qquad i_{W_{\pm}} Z_{\pm} = 0.
\end{equation}
This solves \eqref{eq:0and1formeqsa} and \eqref{eq:0and1formeqsb} without loss of generality. To be able to solve the remaining constraints in \eqref{eq:0and1formeqs}, we decompose $\eta_-$ in terms of $\eta_+$ where, following appendix \ref{app:canonical frame}, $\eta_+$ is our canonical unit norm Majorana spinor
\begin{equation}
\eta_- = a \eta_+ - i b V \eta_+.
\end{equation}
Here $V$ is a unit norm one-form and we take $a^2 + b^2 = 1$. Using this parametrisation, the remaining equations in \eqref{eq:0and1formeqs} can be written as
\begin{subequations}\label{eq:eqsWpmZpm}
\begin{align}
& i_{V} (W_+ + W_-) = i_{V} (Z_+ + Z_-) = 0, \label{eq:eqsWpmZpma} \\[2mm]
& a \left( i_{W_+ - W_-} (Z_+ + Z_-) - i_{W_+ + W_-} (Z_+ - Z_-) \right) + \notag \\[2mm]
& b \left( i_{W_+ - W_-}(i_{Z_+ - Z_-}(i_V \phi_3)) - i_{W_+ + W_-}(i_{Z_+ + Z_-}(i_V \phi_3)) \right) = 0, \label{eq:eqsWpmZpmb} \\[2mm]
& a (W_+ + W_-) + b i_{W_+ - W_-}(i_V \phi_3) = 0, \label{eq:eqsWpmZpmc} \\[2mm]
& a (Z_+ + Z_-) + b i_{Z_+ - Z_-}(i_V \phi_3) = 0, \label{eq:eqsWpmZpmd} \\[2mm]
& a ( i_{Z_-}(i_{W_+}\phi_3 ) + i_{Z_+}(i_{W_-}\phi_3 )) + b ( i_{W_+} (i_{Z_-} ( i_{V} \star_7 \phi_3)) + i_{W_-}(i_{Z_+} (i_V \star_7 \phi_3)) \notag \\[2mm]
& - W_- i_{Z_+} V  + V i_{Z_+} W_- - Z_+ i_{V} W_-  + W_+ i_{Z_-} V - V i_{Z_-} W_+ + Z_- i_{V} W_+ ) = 0, \label{eq:eqsWpmZpme}
\end{align}
\end{subequations}
where $\phi_3$ is the G$_2$-structure three-form associated to $\eta_+$, see \eqref{eq:canonphi3} for its definition. We first observe that $b \neq 0$, otherwise $W_+ = - W_-$ and $Z_+ = - Z_-$, which one can show is inconsistent with the remaining constraints. Given this, in order to further simplify \eqref{eq:eqsWpmZpm}, it is convenient to introduce the following one-forms
\begin{equation}
W_{1,2} = \frac{1}{\sqrt{2}} (W_+ \pm W_-), \qquad Z_{1,2} = \frac{1}{\sqrt{2}} (Z_+ \pm Z_-),
\end{equation}
which unlike $W_{\pm},Z_{\pm}$ are not necessarily unit norm, then \eqref{eq:eqsWpmZpm0} become
\begin{subequations}\label{eq:eqsW12Z120}
\begin{align}
& i_{W_1 \pm W_2}(W_1 \pm W_2) = i_{Z_1 \pm Z_2} (Z_1 \pm Z_2) = 2, \\[2mm]
& i_{W_1 \pm W_2} (Z_1 \pm Z_2) = 0,
\end{align}
\end{subequations}
while \eqref{eq:eqsWpmZpm} can be written as
\begin{subequations}\label{eq:eqsW12Z12}
\begin{align}
& i_{V} W_1 = i_{V} Z_1 = 0, \label{eq:eqsW12Z12a} \\[2mm]
& a \left( i_{W_2} Z_1 - i_{W_1} Z_2 \right) + b \left( i_{W_2}(i_{Z_2}(i_V \phi_3)) - i_{W_1}(i_{Z_1}(i_V \phi_3)) \right) = 0 \label{eq:eqsW12Z12b} \\[2mm]
& a W_1 + b i_{W_2}(i_V \phi_3) = 0, \label{eq:eqsW12Z12c} \\[2mm]
& a Z_1 + b i_{Z_2}(i_V \phi_3) = 0, \label{eq:eqsW12Z12d} \\[2mm]
& a ( i_{W_2}(i_{Z_2}\phi_3 ) - i_{W_1}(i_{Z_1}\phi_3 ) ) + b ( i_{Z_2} (i_{W_2} ( i_{V} \star_7 \phi_3)) - i_{Z_1}(i_{W_1} (i_V \star_7 \phi_3))\notag \\[2mm]
& - W_1 i_{Z_2} V + W_2 i_{Z_1} V - V i_{Z_1} W_2 + V i_{Z_2} W_1 + Z_1 i_V W_2 - Z_2 i_V W_1 ) = 0. \label{eq:eqsW12Z12e}
\end{align}
\end{subequations}
Applying $i_V$ on \eqref{eq:eqsW12Z12e} yields
\begin{equation}
a \left( i_{W_2}(i_{Z_2}(i_V \phi_3)) - i_{W_1}(i_{Z_1}(i_V \phi_3)) \right) + b \left( 2 i_V W_2 i_{Z_1} V - 2 i_V W_1 i_{Z_2} V + i_{W_1} Z_2 - i_{W_2} Z_1 \right) = 0.
\end{equation}
The first two terms in the second round brackets are zero because of \eqref{eq:eqsW12Z12a}, the other terms can be written as
\begin{equation}
a \left( i_{W_2}(i_{Z_2}(i_V \phi_3)) - i_{W_1}(i_{Z_1}(i_V \phi_3)) \right) - b \left( i_{W_2} Z_1 - i_{W_1} Z_2  \right) = 0.
\end{equation}
This equation, combined with \eqref{eq:eqsW12Z12b}, implies that
\begin{subequations} \label{eq:eqsW12Z12new}
\begin{align}
&i_{W_2} Z_1 - i_{W_1} Z_2 = 0, \label{eq:eqsW12Z12newa} \\[2mm]
&i_{W_2}(i_{Z_2}(i_V\phi_3))-i_{W_1}(i_{Z_1}(i_V\phi_3))=0,\label{eq:eqsW12Z12newb}
\end{align}
\end{subequations}
must both be zero. Then \eqref{eq:eqsW12Z120} can be rewritten as
\begin{subequations}\label{eq:eqsW12Z12simp}
\begin{align}
i_{W_1} W_1 + i_{W_2} W_2 & = 2, \qquad i_{W_1} W_2 = 0, \label{eq:eqsW12Z12simpa} \\[2mm] 
i_{Z_1} Z_1 + i_{Z_2} Z_2 & = 2, \qquad i_{Z_1} Z_2 = 0, \label{eq:eqsW12Z12simpb} \\[2mm]
i_{W_1} Z_1 + i_{W_2} Z_2 & = 0, \label{eq:eqsW12Z12simpc} \\[2mm]
i_{W_1} Z_2 + i_{W_2} Z_1 & = 0. \label{eq:eqsW12Z12simpd}
\end{align}
\end{subequations}
By combining \eqref{eq:eqsW12Z12simpd} and \eqref{eq:eqsW12Z12newa} we obtain that
\begin{equation}
i_{W_1} Z_2 = i_{W_2} Z_1 = 0.
\end{equation}
In summary, the zero- and one-form constraints now take the form
\begin{subequations}\label{eq:BPSNeq3intermediate}
\begin{align}
& i_{W_1} W_1 + i_{W_2} W_2 = 2, \\[2mm] 
& i_{Z_1} Z_1 + i_{Z_2} Z_2 = 2, \\[2mm]
& i_{W_1} Z_1 + i_{W_2} Z_2 = 0, \\[2mm] 
& i_{W_1} W_2 = i_{Z_1} Z_2 = i_{W_1} Z_2 = i_{W_2} Z_1 = i_{V} W_1 = i_{V} Z_1 = 0,\\[2mm]
& i_{W_2}(i_{Z_2}(i_V \phi_3)) - i_{W_1}(i_{Z_1}(i_V \phi_3)) = 0, \\[2mm]
& a W_1 + b i_{W_2}(i_V \phi_3) = 0, \\[2mm]
& a Z_1 + b i_{Z_2}(i_V \phi_3) = 0, \\[2mm]
&  b ( i_{Z_2} (i_{W_2} ( i_{V} \star_7 \phi_3)) - i_{Z_1}(i_{W_1} (i_V \star_7 \phi_3)) - W_1 i_{Z_2} V + Z_1 i_V W_2 ) \nn \\[2mm]
& + a ( i_{W_2}(i_{Z_2}\phi_3 ) - i_{W_1}(i_{Z_1}\phi_3 ) ) = 0.
\end{align}
\end{subequations}
One can now explicitly prove that
\begin{equation}\label{eq:WpWmZpZm}
W_+ \neq \pm W_-, \qquad Z_+ \neq \pm Z_-,
\end{equation}
otherwise the equations \eqref{eq:BPSNeq3intermediate} do not admit a solution.
Thanks to \eqref{eq:WpWmZpZm}, we find it convenient to introduce a final basis of unit norm one-forms
\begin{equation}
\begin{aligned}
\hat{W}_1 & = \frac{W_1}{\sqrt{2} \cos \beta_1}, \qquad \hat{W}_2 = \frac{W_2}{\sqrt{2} \sin \beta_1}, \\[2mm]
\hat{Z}_1 & = \frac{Z_1}{\sqrt{2} \cos \beta_2}, \qquad \hat{Z}_2 = \frac{Z_2}{\sqrt{2} \sin \beta_2}.
\end{aligned}
\end{equation}
where
\begin{equation}\label{eq:defbeta12}
\begin{aligned}
\cos \beta_1 & = \sqrt{\frac{1 + i_{W_+} W_-}{2} }, \qquad \sin \beta_1 = \sqrt{\frac{1 - i_{W_+} W_-}{2} }, \\[2mm]
\cos \beta_2 & = \sqrt{\frac{1 + i_{Z_+} Z_-}{2} }, \qquad \sin \beta_2 = \sqrt{\frac{1 - i_{Z_+} Z_-}{2} },
\end{aligned}
\end{equation}
with $\cos \beta_1 \neq 0$ and $\sin \beta_1 \neq 0$, as a consequence of \eqref{eq:WpWmZpZm}\footnote{Furthermore, it is clear from \eqref{eq:defbeta12} that $0 < \beta_1 < \frac{\pi}{2}$ and $0 < \beta_2 < \frac{\pi}{2}$.}. In this new parametrisation, the conditions become
\begin{subequations}\label{eq:finaleqNeq3}
\begin{align}
& i_{\hat{W}_1} \hat{W}_1 = i_{\hat{W}_2} \hat{W}_2 = i_{\hat{Z}_1} \hat{Z}_1 = i_{\hat{Z}_2} \hat{Z}_2 = 1, \label{eq:finaleqNeq3a} \\[2mm] 
& i_{\hat{W}_1} \hat{W}_2 = i_{\hat{Z}_1} \hat{Z}_2 = i_{V} \hat{W}_1 = i_{V} \hat{Z}_1 = i_{\hat{W}_1} \hat{Z}_2 = i_{\hat{W}_2} \hat{Z}_1 = 0, \label{eq:finaleqNeq3b} \\[2mm]
& \cos \beta_2 \cos \beta_1 i_{\hat{W}_1} \hat{Z}_1 + \sin \beta_2 \sin \beta_1 i_{\hat{W}_2} \hat{Z}_2 = 0, \label{eq:finaleqNeq3c} \\[2mm]
& \sin \beta_2 \sin \beta_1 i_{\hat{W}_2}(i_{\hat{Z}_2}(i_V \phi_3)) - \cos \beta_2 \cos \beta_1 i_{\hat{W}_1}(i_{\hat{Z}_1}(i_V \phi_3)) = 0, \label{eq:finaleqNeq3d} \\[2mm]
& a \cos \beta_1 \hat{W}_1 + b \sin \beta_1 i_{\hat{W}_2}(i_V \phi_3) = 0, \label{eq:finaleqNeq3e} \\[2mm]
& a \cos \beta_2 \hat{Z}_1 + b \sin \beta_2 i_{\hat{Z}_2}(i_V \phi_3) = 0, \label{eq:finaleqNeq3f} \\[2mm]
& a ( \sin \beta_2 \sin \beta_1 i_{\hat{W}_2}(i_{\hat{Z}_2}\phi_3 ) - \cos \beta_2 \cos \beta_1 i_{\hat{W}_1}(i_{\hat{Z}_1}\phi_3 ) ) + b ( \sin \beta_2 \sin \beta_1 i_{\hat{Z}_2} (i_{\hat{W}_2} ( i_{V} \star_7 \phi_3)) \notag \\[2mm]
& - \cos \beta_2 \cos \beta_1 i_{\hat{Z}_1}(i_{\hat{W}_1} (i_V \star_7 \phi_3)) - \cos \beta_1 \sin \beta_2 \hat{W}_1 i_{\hat{Z}_2} V + \cos \beta_2 \sin \beta_1 \hat{Z}_1 i_{\hat{W}_2} V ) = 0. \label{eq:finaleqNeq3g}
\end{align}
\end{subequations}
At this point we proceed to solve \eqref{eq:finaleqNeq3}. Without loss of generality we can expand the one-forms $(V, \hat{W}_{1,2}, \hat{Z}_{1,2})$ in the following basis of vielbein elements on M$_7$;
\begin{equation}
\begin{aligned}
V & = v^5 e^5 + v^6 e^6 + v^7 e^7, \qquad \hat{W}_1 = w^7_1 e^7, \qquad \hat{W}_2 = w^6_2 e^6 + w^7_2 e^7, \\[2mm]
\hat{Z}_1 & = z_1^4 e^4 + z_1^5 e^5 + z_1^6 e^6 + z_1^7 e^7, \qquad \hat{Z}_2 = z_2^3 e^3 + z_2^4 e^4 + z_2^5 e^5 + z_2^6 e^6 + z_2^7 e^7,
\end{aligned}
\end{equation}
where the coefficients are all functions that depend on the internal manifold M$_7$. Substituting these into \eqref{eq:finaleqNeq3}, we find two branches depending on whether $a$ vanishes or not.

In the case $a=0$, without loss of generality, the conditions \eqref{eq:finaleqNeq3} are solved by
\begin{equation}
V = e^6, \qquad \hat{W}_1 = e^7, \qquad \hat{W}_2 = e^6 \qquad \hat{Z}_1 = e^7, \qquad \hat{Z}_2 =  e^6,
\end{equation}
which solve all the constraints except for two, namely \eqref{eq:finaleqNeq3c} and \eqref{eq:finaleqNeq3g}, which reduce to
\begin{equation}\label{eq:nogofirstbranching}
\begin{aligned}
\eqref{eq:finaleqNeq3c} \to \cos (\beta_1 - \beta_2) & = 0, \\[2mm]
\eqref{eq:finaleqNeq3g} \to \sin( \beta_1 - \beta_2) & = 0.
\end{aligned}
\end{equation}
It is clear that \eqref{eq:nogofirstbranching} cannot be solved simultaneously, so this branch cannot be solved.

For the other branch, we must take $a \neq 0$. Also in this case, one can solve all the constraints except for \eqref{eq:finaleqNeq3f} and \eqref{eq:finaleqNeq3g} by fixing
\begin{equation}
\begin{aligned}
V & = \frac{- a \cot \beta_1 e^5 - \sqrt{b^2 - a^2 \cot^2 \beta_1} e^6}{b}, \qquad \hat{W}_1 = e^7, \qquad \hat{W}_2 = e^6 \qquad \hat{Z}_1 = e^7, \\[2mm]
\hat{Z}_2 & =  \sqrt{1 - \cot^2 \beta_1 \cot^2 \beta_2} e^5 - \cot \beta_1 \cot \beta_2 e^6,
\end{aligned}
\end{equation}
where we remind the reader that $\cos \beta_i \neq 0$ and $\sin \beta_i \neq 0$. One then finds that \eqref{eq:finaleqNeq3f} and \eqref{eq:finaleqNeq3g} become
\begin{equation}\label{eq:nogosecondbranching}
\begin{aligned}
\eqref{eq:finaleqNeq3f} \to a \frac{\cos \beta_2}{\sin^2 \beta_1} - \sin \beta_2 \sqrt{b^2 -a^2 \cot^2 \beta_1}\sqrt{1 - \cot^2 \beta_1 \cot^2 \beta_2 } = 0 , \\[2mm]
\eqref{eq:finaleqNeq3g} \to \cos \beta_2 \sqrt{b^2 - a^2 \cot^2 \beta_1} + a \sin \beta_2 \sqrt{1 - \cot^2 \beta_1 \cot^2 \beta_2} = 0.
\end{aligned}
\end{equation}
One can check that the only way to solve these is to set $a=b=0$, which is incompatible with $a^2 + b^2 = 1$, and so the second branch suffers a similar fate to the first.

We thus have shown that the only way to realise ${\cal{N}} = 3$ is to have one of $\lambda_{\pm} = 0$, so the Tri-Sasaki limit that we have derived in section \ref{subsec:TriSasaki}.

\section{Conclusions}

This work focuses on the classification of $ {\cal N}\leq 3$ supersymmetric AdS$_4$ solutions of $d=11$ supergravity. This is achieved with a combination of G-structure and more standard spinorial techniques that allow supersymmetry to be expressed geometrically in terms of differential constraints on the forms spanning the G-structure in question.

The main new results of this work are two fold. 1) A proof that all $ {\cal N}= 3$ supersymmetric AdS$_4$ solutions must have an internal space that is Tri-Sasaki. It follows that the same must be true of any solution preserving $ {\cal N}\geq 3$. 2) A derivation of the necessary and sufficient conditions for totally general AdS$_4$ solutions in $d=11$ to preserve ${\cal N}=1$ supersymmetry. These conditions are purely geometric and are expressed in terms of differential constraints on the forms that span the SU(3)-structure that the internal space of such a solution generically supports.
~\\

Some interesting future directions are the following:

\begin{itemize}
\item It would be interesting to address the statement of ${\cal{N}} = 3 $ AdS$_4$ solutions both in massive type IIA and type IIB. There exists an ${\cal{N}}=3$ AdS$_4$ vacuum \cite{DeLuca:2018buk,Pang:2015vna,Pang:2015rwd} in massive type IIA whose internal manifold is not related to any Tri-Sasaki manifold. On the other hand, in type IIB, a well-known family of ${\cal{N}}=4$ AdS$_4 \times \text{S}^2 \times \text{S}^2 \times \Sigma_2$ solutions has long been known \cite{DHoker:2007zhm,DHoker:2007hhe}. One would expect a deformation of this class such that one S$^2$ becomes fibred over the other S$^2$, breaking the SO($4$) R-symmetry to SO($3$). Furthermore, the type IIA solution of \cite{DeLuca:2018buk,Pang:2015vna,Pang:2015rwd} has an SO(4) isometry manifest in the metric, which decomposes into a flavour and R symmetry symmetry SO(3), $ \text{SO}(4)= \text{SO}(3)_R \times \text{SO}(3)_F$. Given that SO(3)$_F$ is realised by SU(2) Maurer Cartan forms, and that the spinors of the geometry are SO(3)$_F$ singlets, one can apply either an abelian or a non-Abelian T-duality on this $\text{SO}(3)_F$ arriving in type IIB without breaking supersymmetry. This provides two distinct ${\cal{N}}=3$ solutions, that suggest a more general class. Such observations have in the past lead to the construction of new and interesting classes of solutions such as $\text{AdS}_2 \times \mathbb{CP}^3 \times \Sigma_2 $ \cite{Conti:2025djz} and $\text{AdS}_3 \times \text{S}^2 \times \text{CY}_2 \times I$ \cite{Lozano:2019ywa,Lozano:2019emq,Lozano:2019zvg}, in these instances the first geometries suggestive of the broader classes were derived in \cite{Conti:2023rul,Cho:2002zp,Sfetsos:2010uq,Lozano:2019ywa} by operating with (non)-Abelian T-duality on the $\text{AdS}_3 \times \mathbb{CP}^3 \times I $ solutions of \cite{Macpherson:2023cbl} and the $\text{AdS}_3 \times \text{S}^3 \times \mathbb{T}^4$ solutions of \cite{Tseytlin:1996bh,Maldacena:1997re}. All this suggests that a focused classification effort in type IIB will be productive.
\item One could investigate the existence of ${\cal{N}}=3$ solutions for AdS spacetimes of other dimensions. It is known that ${\cal{N}}=3$ supersymmetric AdS$_5$ solutions are locally AdS$_5 \times \text{S}^5$ \cite{Beck:2016lwk}, where a concrete example was found in \cite{Garcia-Etxebarria:2015wns}. For AdS$_3$, solutions preserving ${\cal{N}}=(3,0)$ \cite{Legramandi:2019xqd} and ${\cal{N}}=(3,3)$ \cite{Eberhardt:2018sce} are known, however it is likely that many more exist and a systematic study of what can be constructed is currently lacking.
\item It is known that AdS$_4$ solutions in eleven dimensions with ${\cal{N}} > 4$ are locally AdS$_4 \times \text{S}^7 / {\Gamma}$ \cite{Haupt:2017bnj}. In this paper we classified AdS$_4$ solutions with ${\cal{N}} \leq 3$ and showed that all solutions preserving ${\cal{N}} \geq 3$ supersymmetry necessarily admit a Tri-Sasaki structure. A natural question is which internal spaces M$_7$ can support ${\cal{N}}=4$ supersymmetry, apart from quotients of the seven-sphere. A qualitative argument goes as follows. Since we know M$_7$ must admit a Tri-Sasaki structure, and the solution is of Freund-Rubin type, ${\cal{N}}=4$ would require M$_7$ to admit four Killing spinors. However, the only seven-dimensional manifolds admitting a Tri-Sasaki structure and four Killing spinors are the round seven-sphere and its quotients, see the discussion following theorem 4.4 of \cite{Agricola:2006tx}. We leave a more systematic investigation of this question for future work.
\end{itemize}
It would be interesting to explore some of these directions in the future.

\section*{Acknowledgements}

I thank Niall Macpherson for insightful discussions during this project and Chris Couzens for useful conversations. I also thank Achilleas Passias and Yolanda Lozano for comments on the draft. My work is partially supported by the grants from the Spanish government MCIU-22-PID2021-123021NB-I00 and MCIU-25-PID2024-161500NB-I00. My work is funded by the Severo Ochoa fellowship PA-23-BP22-019.

\appendix

\section{Conventions}\label{sec:Conventions}

The bosonic fields of eleven dimensional supergravity consist of a metric $g$, and a three-form potential $C$ with four-form field strength $G = dC$. The action for the bosonic sector is
\begin{equation}\label{eq:action11D}
S= \frac{1}{2 k^2 } \int R \star 1 - \frac{1}{2} G \wedge \star G - \frac{1}{6}G \wedge G \wedge C.
\end{equation}
The equations of motion that follow from \eqref{eq:action11D} are
\begin{subequations}\label{eq:eom11D}
\begin{align}
& R_{M N} - \frac{1}{12} \left( G_{M A B C} \tensor{G}{_N ^A ^B ^C} - \frac{1}{12} g_{M N} G_{ABCD} G^{ABCD} \right) = 0, \label{eq:eom11DEinstein} \\[2mm]
& d \star G + \frac{1}{2} G \wedge G = 0, \label{eq:eom11Dflux}
\end{align}
\end{subequations}
and the Bianchi identity is
\begin{align}\label{eq:Bianchi11D}
d G = 0.
\end{align}
Our convention for the hodge dual is
\beq
\star e^{\underline{M}_1...\underline{M}_k}=\frac{1}{(d-k)!}\epsilon_{\underline{M}_{k+1}...\underline{M}_{d-k}}^{~~~~~~~~~~~~~\underline{M}_1...\underline{M}_k}e^{\underline{M}_{k+1}...\underline{M}_{d-k}},
\eeq
where the $d=11$ indices $M$ are curved and $\underline{M}$ are flat. The Killing spinor equation is
\begin{align}\label{eq:KSE11D}
\nabla_{M} \epsilon + \frac{1}{288} \left( \tensor{\Gamma}{_M ^{ABCD}} - 8 \delta_M ^A \Gamma^{BCD} \right) G_{ABCD} \epsilon = 0.
\end{align}
An equivalent expression, with which the reader may be more familiar, is the following
\begin{align}
\nabla_{M} \epsilon + \frac{1}{24} \left( 3 \slashed{G} \Gamma_M - \Gamma_M \slashed{G} \right) \epsilon = 0.
\end{align}

\section{Derivation of ${\cal{N}}=1$ AdS$_4$ conditions}\label{app:Neq1}

Although systems preserving ${\cal{N}}=1$ supersymmetry have been derived before in \cite{Lukas:2004ip,Behrndt:2005im}, \cite{Gabella:2012rc} pointed out these were not fully general, so we derive them independently. We achieve this by taking the AdS$_3$ conditions from \cite{Legramandi:2020txf}, which are necessary and sufficient, and impose an enhancement to AdS$_4$.

\subsection{Review of AdS$_3$ conditions}

We begin by reviewing the AdS$_3$ conditions for ${\cal{N}}=1$ supersymmetry, following the derivation presented in \cite{Legramandi:2020txf}. We decompose the bosonic fields as
\begin{equation}
ds^2 = e^{2 \Delta} ds^2(\text{AdS}_3) + ds^2(\text{M}_8), \qquad G = e^{3 \Delta} \text{vol} (\text{AdS}_3) \wedge G_1 + G_4,
\end{equation}
and the eleven dimensional Killing spinor as
\begin{equation}
\epsilon = \zeta \otimes \eta^{(8)},
\end{equation}
where $\zeta$ is the AdS$_3$ Killing spinor satisfying
\begin{equation}
\nabla_{\mu}^{(3)} \zeta = \frac{m}{2} \gamma_{\mu}^{(3)} \zeta,
\end{equation}
and $\eta^{(8)}$ a non chiral spinor on M$_8$. Both $\zeta$ and $\eta^{(8)}$ are Majorana. We work in conventions such that
\begin{equation}
\Gamma_{\mu} = \gamma_{\mu}^{(3)} \otimes \hat{\gamma}_8, \qquad \Gamma_{a} = \mathbb{I}_2 \otimes \gamma_a ,
\end{equation}
with
\begin{equation}
\hat{\gamma}_8 = \gamma_{12345678}, \qquad \Gamma_{012345678910} = \mathbb{I}, \qquad B_{11} = \mathbb{I}_2 \otimes B_8.
\end{equation}
If we define the following bilinears,
\begin{equation}\label{eq:8dbilinears}
2 e^A f = \eta^{(8)\dagger} \hat{\gamma}_8 \eta^{(8)}, \qquad 2 e^A K_a^{(8)} = \eta^{(8) \dagger} \gamma_a \eta^{(8)}, \qquad 2 e^A (\Psi_3)_{abc} = \eta^{(8) \dagger} \gamma_{abc} \hat{\gamma}_8 \eta^{(8)},
\end{equation} 
it was shown in \cite{Legramandi:2020txf} that $d=11$ supersymmetry demands the following eight dimensional conditions to be solved which, in our conventions for the hodge dual, are
\begin{subequations}\label{eq:AdS3susy}
\begin{align} 
& d(e^{2 \Delta} K^{(8)})=0 \label{eq:AdS3susy1},\\[2mm]
&d(e^{3 \Delta} f)- e^{3\Delta} G_1-2 m  e^{2\Delta} K^{(8)} =0, \label{eq:AdS3susy2}\\[2mm]
& d(e^{3\Delta} \Psi_3)- e^{3\Delta}(-\star_8 G_4+ f G_4)+2 m e^{2\Delta} \Psi_4=0, \label{eq:AdS3susy3}\\[2mm]
&d (e^{2\Delta} \Psi_4)- e^{2\Delta} K^{(8)} \wedge G_4=0, \label{eq:AdS3susy4}\\[2mm]
&6 \star_8 d\Delta - 2 f \star_8 G_1 - \Psi_3\wedge G_4=0, \label{eq:AdS3susy5}\\[2mm]
&6m \text{Vol}(\text{M}_8)+ e^{\Delta}(- 2 K^{(8)} \wedge \star_8 G_1+ \Psi_4\wedge G_4)=0\label{eq:AdS3susy6},
\end{align}
\end{subequations}
These conditions are necessary but not sufficient, an additional condition is required for sufficiency
\begin{equation}\label{eq:AdS3susysuff}
6 e^{- \Delta} m \star_8 K^{(8)} - 6 f \star_8 d \Delta + 2 \star_8 G_1 - \Psi_3 \wedge \star_8 G_4 = 0.
\end{equation}

\subsection{Sufficient conditions for ${\cal{N}}=1$ AdS$_4$ }

We now derive the ${\cal{N}}=1$ AdS$_4$ conditions starting from these. We begin by decomposing
\begin{equation}
e^{\Delta} = e^A \cosh{(m x)}, \qquad e^{3 \Delta} G_1 = c \cosh{(m x)} dx, \qquad ds^2(\text{M}_8) = dx^2 + ds^2(\text{M}_7),
\end{equation}
in this way we make the AdS$_4$ factor in the bosonic fields manifest
\begin{equation}
ds^2 = e^{2A} ds^2(\text{AdS}_4) + ds^2(\text{M}_7), \qquad G = c \, \text{vol}(\text{AdS}_4) + G_4,
\end{equation}
where
\begin{equation}
ds^2(\text{AdS}_4) = \cosh{(2 m x)} ds^2(\text{AdS}_3) + dx^2.
\end{equation}
For the spinor, we perform the following $1 + 7$ split
\begin{equation}\label{eq:17split}
\gamma_1 = \sigma_1 \otimes \mathbb{I}_8, \qquad \gamma_{a+1} = \sigma_2 \otimes \gamma_{a}, \qquad B_8 = \mathbb{I}_2 \otimes B_7,
\end{equation}
for $a = {1,..,7}$ and such that $i \gamma_{1234567} = 1$. We then write the most general seven dimensional Dirac spinor as 
\begin{equation}
\chi = \sqrt{e^A} \left( \chi_1 + i \chi_2 \right),
\end{equation}
where $\chi_{1,2}$ are unit norm Majorana spinors. We introduce the following two-dimensional vector
\begin{align} 
\theta_+ = 
	\begin{pmatrix}
	1 \\
	0
	\end{pmatrix}, \qquad
\theta_- & =
	\begin{pmatrix}
	0 \\
	1
	\end{pmatrix}.
\end{align}
The most general eight dimensional spinor $\eta^{(8)}$, such that undergoes the $1+7$ split in \eqref{eq:17split}, is
\begin{equation}
\eta^{(8)} = \theta_+ \otimes \left( \cosh \left( \frac{m x}{2} \right) \chi_1  + \sinh \left( \frac{m x}{2}\right) \chi_2 \right) + \theta_- \otimes \left( \sinh \left( \frac{m x}{2} \right) \chi_1 - \cosh \left( \frac{m x}{2}\right) \chi_2 \right),
\end{equation}
and such that the Killing spinors of plus and minus chirality on AdS$_4$ are
\begin{equation}
\zeta_{\pm} =  \frac{1}{2} \left( \cosh \left( \frac{m x}{2} \right) - i \sinh \left( \frac{m x}{2} \right) \right) \zeta \otimes v_{\pm},
\end{equation}
where
\begin{align} 
v_{\pm} = 
	\begin{pmatrix}
	1 \\
	\pm i
	\end{pmatrix}.
\end{align}
In this way, with the $3+1$ split given by
\begin{equation}
\gamma^{(4)}_{\mu} = \gamma^{(3)}_{\mu} \otimes \sigma_3, \qquad \gamma_4^{(4)} = \mathbb{I}_2 \otimes \sigma_1,
\end{equation}
and $\hat{\gamma}^{(4)} = - i \gamma_{1234}$. $\zeta_{\pm}$ solve the AdS$_4$ Killing spinor equation
\begin{equation}
\nabla^{(4)}_{\mu} \zeta_{\pm} = \frac{m}{2} \gamma^{(4)}_{\mu} \zeta_{\mp}.
\end{equation}
The eleven dimensional Killing spinor can be written as
\begin{equation}
\epsilon = \zeta_+ \otimes \chi + \text{m.c.} ,
\end{equation}
where m.c. denotes the Majorana conjugate. The seven dimensional spinor defines the following bilinears
\begin{equation}\label{eq:defXNeq1}
\chi^{\dagger}\chi = e^A, \qquad X_n = \frac{1}{n!} \chi^{\dagger} \gamma_{a_1 ... a_n} \chi e^{a_1... a_n},
\end{equation}
and using the Majorana conjugate we can also build
\begin{equation}\label{eq:defYNeq1}
Y_n = \frac{1}{n!} \chi^{c \dagger} \gamma_{a_1 ... a_n} \chi e^{a_1 ... a_n}.
\end{equation}
In terms of these, we can rewrite the eight dimensional bilinears \eqref{eq:8dbilinears} as
\begin{align}\label{eq:8dbilto7dbil}
e^A K^{(8)} & = \text{Re} X_1 + ( \text{Re}Y_0 \tanh(mx) - \sech(mx) \text{Im}Y_0 ) dx, \nn \\[2mm]
e^A \Psi_3 & = \text{Im}X_2 \wedge dx - \sech(mx) \text{Re}Y_3 -\tanh(mx) \text{Im} Y_3, \\[2mm]
e^A \Psi_4 & = \text{Re}X_4 + dx \wedge ( \tanh(mx) \text{Re} Y_3 - \sech(mx) \text{Im}Y_3 ) . \nn
\end{align}
By plugging this $3 + 1 + 7 $ split into \eqref{eq:AdS3susy} we obtain
\begin{subequations}\label{eq:AdS4N=1}
\begin{align}
& 3 e^{2A} m \text{Im}Y_0 - c = 0,  \label{eq:AdS4N=1a} \\[2mm]
& d \left( e^{2A} \text{Im}Y_0 \right) = 0, \label{eq:AdS4N=1b} \\[2mm]
& d \left( e^{2A} \text{Re}Y_0 \right) - 2 m e^{A} \text{Re}X_1 = 0, \label{eq:AdS4N=1c} \\[2mm]
& d \left( e^{A} \text{Re}X_1 \right) = 0, \label{eq:AdS4N=1d} \\[2mm]
& 6 e^A \star_7 dA - \text{Im}X_2 \wedge G_4 = 0, \label{eq:AdS4N=1e} \\[2mm]
& (6 m e^{4A} - 2 c \text{Im}Y_0 ) \text{vol}_7 - e^{4A} \text{Im}Y_3 \wedge G_4 = 0, \label{eq:AdS4N=1f} \\[2mm]
& 2 \text{Re}Y_0 c \text{vol}_7 + e^{4A} \text{Re}Y_3 \wedge G_4 = 0, \label{eq:AdS4N=1g} \\[2mm]
& d \left( e^{3 A} \text{Im}X_2 \right) + 3 m e^{2A} \text{Im}Y_3 - e^{4A} \star_7 G_4 = 0, \label{eq:AdS4N=1h} \\[2mm]
& d \left( e^{2 A} \text{Im}Y_3 \right) + e^{2A} \text{Im}Y_0 G_4 = 0, \label{eq:AdS4N=1i} \\[2mm]
& d \left( e^{2 A} \text{Re}Y_3 \right) - 2 m e^A \text{Re}X_4 + e^{2A} \text{Re}Y_0 G_4 = 0, \label{eq:AdS4N=1j} \\[2mm]
& d \left( e^A \text{Re}X_4 \right) - e^A \text{Re}X_1 \wedge G_4 = 0. \label{eq:AdS4N=1k}
\end{align}
\end{subequations}
The conditions \eqref{eq:AdS4N=1} a priori are necessary but not sufficient, to have sufficient conditions we need to check explicitly \eqref{eq:AdS3susysuff}. However, upon inserting \eqref{eq:8dbilto7dbil} into \eqref{eq:AdS3susysuff} we find that it is implied by \eqref{eq:AdS4N=1}.
~\\

We conclude this appendix with a few observations on the $a_i$ appearing in \eqref{eq:bilinearsparamNeq1}, by checking their consistency with \eqref{eq:AdS4N=1}. First we observe that \eqref{eq:AdS4N=1a} and \eqref{eq:AdS4N=1b} imply $c$ is a constant.

In general, we note that setting $a_1 = 0$ implies $c=0$ from \eqref{eq:AdS4N=1a}, in turn \eqref{eq:AdS4N=1e} implies $A$ is constant. Looking at \eqref{eq:AdS4N=1j} and \eqref{eq:AdS4N=1h}, we see the conditions for $G_4$ and $\star_7 G_4$ are incompatible. This fixes $G_4 = m = 0$ that kills the AdS$_4$ class. Focusing on the case $a_2 = 0$, this still implies $c=0$ because of \eqref{eq:AdS4N=1a}, allowing for purely magnetic four-form flux. The case $a_3=0$, similarly to the first case, implies $A$ to be constant \eqref{eq:AdS4N=1e}, while \eqref{eq:AdS4N=1b} and \eqref{eq:AdS4N=1c} set $a_1$ and $a_2$ to be constants. At this stage we take the linear combination 
\begin{equation}
2 a_1 a_2 \eqref{eq:AdS4N=1i} - (a_1^2 - a_2^2) \eqref{eq:AdS4N=1j} = e^{2A} \left(m \left(a_1^2 - a_2^2 \right) (J\wedge J+2 \text{Re} \Omega \wedge V) + e^{A} G_4 \right).
\end{equation}
This expression for $G_4$, together with \eqref{eq:AdS4N=1h}, can be simultaneously solved only if $a_1 = \pm a_2$ and again $G_4 = 0$. We note that, in this case, we do not need to set $m = 0$, so the class does not die, indeed this spinor defines a weak G$_2$-structure.
~\\

We observe that in terms of bilinears $a_1 = 0$ is equivalent to demanding $\text{Re} Y_0 = - e^A$, $\text{Im} Y_0 = 0$ and $X_1 = X_2 = 0$. The condition $a_2 = 0$ implies $\text{Im}Y_0 = 0$, $\text{Re} Y_0 \neq - e^A$ and $X_1 \neq 0$ with $X_2 \neq 0$. The case $a_3 = 0$ is equivalent to $\text{Re}Y_0 \neq - e^{A}$ and $\text{Im} Y_0 \neq 0$ such that $ \text{Re}Y_0^2 + \text{Im}Y_0^2 = e^{2A}$, with moreover $X_1 = X_2 = 0$.

\section{${\cal{N}}=3$ AdS$_4$} \label{app:Neq3}

The eleven dimensional supersymmetry conditions have been worked out in \cite{Gauntlett:2003wb,Gauntlett:2002fz}. By defining the following bilinears
\begin{equation}\label{eq:bilinears11D}
K_M = \bar{\epsilon} \Gamma_M \epsilon, \qquad \Xi_{MN} = \bar{\epsilon} \Gamma_{MN} \epsilon, \qquad \Sigma_{MNOPQ}= \bar{\epsilon} \Gamma_{MNOPQ} \epsilon,
\end{equation}
they can be written as
\begin{subequations}\label{eq:BPS11}
\begin{align}
\nabla_{(M} K_{N)} & = 0, \label{eq:BPS11a}\\[2mm]
\star_{11} d K & = \frac{2}{3} \Xi \wedge \star_{11} G - \frac{1}{3} \Sigma \wedge G, \label{eq:BPS11b} \\[2mm]
d \Xi & = K \lrcorner G, \label{eq:BPS11c} \\[2mm]
d \Sigma & = K \lrcorner \star_{11} G - \Xi \wedge G. \label{eq:BPS11d}
\end{align}
\end{subequations}
These conditions are in general necessary but not sufficient when $K$ is a null Killing vector. However, for AdS$_3$, \eqref{eq:AdS3susy} follows exclusively from these constraints. As established in the previous section, \eqref{eq:AdS3susy} yields necessary and sufficient conditions for AdS$_4$; therefore, for AdS$_4$ in particular, \eqref{eq:BPS11} does provide necessary and sufficient conditions for supersymmetry.

\subsection{Spinors and bilinears on AdS$_4$}

We decide to parametrise AdS$_4$ as
\begin{equation}
ds^2({\text{AdS}_4}) = e^{2 m r} (- dt^2 + dx_1^2 + dx_2^2) + dr^2,
\end{equation}
here $m$ is the inverse radius of AdS$_4$, $R_{\mu \nu}^{(4)} = - 3 m^2 g_{\mu \nu}^{(4)}$ for $g_{\mu \nu}^{(4)}$ where $\mu = (t,x_1, x_2, r)$. We use the obvious vielbein
\begin{equation}
e^{\underline{\mu}} = (e^{m r} dt, e^{m r} dx_1, e^{m r} dx_2, dr ) ,
\end{equation}
where $\underline{\mu}= (1,2,3,4)$. We choose conventions such that we have real flat gamma matrices $\gamma_{\underline{\mu}}^{(4)}$ and we take the chirality matrix to be $ \hat{\gamma}^{(4)} = - i \gamma_{0123} $. The curved space counterparts are obtained by $\gamma_{\mu} = \tensor{e}{^{\underline{\mu}}_{\mu}} \gamma_{\underline{\mu}} $. The Killing spinor equation on AdS$_4$ is
\begin{equation}\label{eq:KSEAdS4}
\nabla_{\mu}^{(4)} \zeta = \frac{m}{2} \gamma_{\mu}^{(4)} \zeta,
\end{equation}
where $\zeta$ is a Killing spinor on AdS$_4$. We take it to be Majorana, that in our conventions is equivalent to take it to be real. Equation \eqref{eq:KSEAdS4} is solved by the spinor
\begin{equation}
\zeta = e^{\frac{m}{2} r } \zeta_0 + \left( e^{- \frac{m}{2} r } + m e^{ \frac{m}{2} r} \left( t \gamma_0^{(4)} + x_1 \gamma_1^{(4)} + x_2 \gamma_2^{(4)} \right)  \right) \tilde{\zeta}_0.
\end{equation}
$\zeta_0$ and $\tilde{\zeta}_0$ are eigenvectors of $\gamma_3^{(4)}$ with eigenvalue $\pm 1$ respectively. These contain the Poincar\'e and superconformal charges respectively. For our purpose, we want to define a triplet of spinors. It is more convenient to determine a basis of gamma matrices
\begin{equation}
\gamma_{0}^{(4)} = i \sigma_2 \otimes \mathbb{I}_2, \qquad \gamma_{1}^{(4)} = \sigma_1 \otimes \mathbb{I}_2, \qquad \gamma_{2}^{(4)} = \sigma_3 \otimes \sigma_1, \qquad \gamma_{3}^{(4)} = \sigma_3 \otimes \sigma_3.
\end{equation}
Given this basis, we define the following Killing spinors 
\begin{equation}
\begin{aligned}
\zeta_1 &= e^{\frac{m}{2}r}
\begin{pmatrix}
1 \\ 0 \\ 0 \\ 0
\end{pmatrix}, \qquad
\zeta_2 = \left( e^{-\frac{m}{2}r} + m e^{\frac{m}{2}r}
\left( t \gamma_0^{(4)} + x_1 \gamma_1^{(4)} + x_2 \gamma_2^{(4)} \right) \right)
\begin{pmatrix}
0 \\ 0 \\ 1 \\ 0
\end{pmatrix}, \\[2mm]
\zeta_3 &= e^{\frac{m}{2}r}
\begin{pmatrix}
0 \\ 0 \\ 0 \\ 1
\end{pmatrix}, \qquad
\zeta_4 = \left( e^{-\frac{m}{2}r} + m e^{\frac{m}{2}r}
\left( t \gamma_0^{(4)} + x_1 \gamma_1^{(4)} + x_2 \gamma_2^{(4)} \right) \right)
\begin{pmatrix}
0 \\ 1 \\ 0 \\ 0
\end{pmatrix}.
\end{aligned}
\end{equation}
We work in conventions such $\bar{\zeta} = (\gamma_{0}^{(4)} \zeta)^{\dagger}$, and the only zero-form bilinears we can define without using the chiral matrix (and that are not trivial) are
\begin{equation}
\bar{\zeta_1} \zeta_2 = - \bar{\zeta}_3 \zeta_4 = - 1.
\end{equation}
For the purpose of this article, we are interested in defining the following triplet of spinors
\begin{equation}
\zeta^{I} = 
\begin{pmatrix}
- c_1 \zeta_1 \\ c_2 \zeta_2 \\ c_3( \zeta_1 + \zeta_2 + \zeta_3)
\end{pmatrix}.
\end{equation}
The $c_I$ are constants and one can check that
\begin{equation}\label{eq:defcalS}
\bar{\zeta}^I \zeta^J = \Omega^{IJ} = \sum_{K=1}^{3} \sum_{L=K}^{3} c^L c^K ({\cal{S}}^{LK})^{IJ}, \qquad ({\cal{S}}^{LK})^{IJ} = (\delta^{IK}\delta^{JL} - \delta^{JK}\delta^{IL}),
\end{equation}
where ${\cal{S}}^{LK}$ span the Lie algebra of $\mathrm{SO}(3)$, and with our choice of $\zeta^I$ each independent $\mathrm{SO}(3)$ generator is required to span $\Omega^{IJ}$.
In a similar fashion, we can define the one-forms and the real zero-forms as
\begin{equation}
f^{IJ} = i\,\bar{\zeta}^I \hat{\gamma}^{(4)} \zeta^J, \qquad
k^{IJ} = \bar{\zeta}^I \gamma_\mu^{(4)} \zeta^J\, dx^\mu, \qquad
u^{IJ} = -i\,\bar{\zeta}^I \gamma_\mu^{(4)} \hat{\gamma}^{(4)} \zeta^J\, dx^\mu,
\end{equation}
where $(f^{IJ}, u^{IJ})$ are antisymmetric and satisfy
\begin{equation}
df^{IJ} = m\, u^{IJ}, \qquad
\nabla_\mu u_\nu^{IJ} = m f^{IJ} g_{\mu\nu}^{(4)}.
\end{equation}
Furthermore, the $f^{IJ}$ are related to the AdS$_4$ embedding coordinates. Indeed, given
\begin{equation}
\begin{aligned}
X_0 & = \frac{e^{-mr}}{2} \left( 1 + e^{2mr}
\left(\frac{1}{m^2} + x_1^2 + x_2^2 - t^2 \right) \right), \qquad X_1 = \frac{e^{mr}}{m} x_1, \qquad X_2 = \frac{e^{mr}}{m} x_2, \\[2mm]
X_3 & = \frac{e^{-mr}}{2} \left( 1 + e^{2mr}
\left(\frac{1}{m^2} + x_1^2 + x_2^2 - t^2 \right) \right), \qquad X_4 = \frac{e^{mr}}{m} t, \\[2mm]
\end{aligned}
\end{equation}
it follows that
\begin{align}
f^{12} &= - m^2 c_1 c_2 X_2, \qquad
f^{13} = - m^2 c_1 c_3 (X_2 - X_0 + X_3), \qquad
f^{23} =-  m^2 c_2 c_3 (X_1 + X_2 + X_4).
\end{align}
$k^{IJ}$ is symmetric, therefore it has 6 components, and one can show that
\begin{equation}
\nabla_{(\mu} k_{\nu)}^{IJ} = 0,
\end{equation}
making $k^{IJ}$ the one-form dual to the Killing vector, while $u^{IJ}$ the one-form dual to the conformal Killing vector.

To know how the triplet of spinors transforms with respect to the Killing vector $k_{\mu}^{IJ}$ we compute the Lie derivative
\begin{equation}\label{eq:LiederivspinorAdS4}
{\cal{L}}_{k^{IJ}} \zeta^K = 2 m (\Lambda^{IJ})^{KL} \zeta^L,
\end{equation}
where $\Lambda^{IJ}$ are symmetric matrices
\begin{equation}
\begin{aligned}
\Lambda^{11} &=
\begin{pmatrix}
0 & 0 & 0\\
c_1 c_2 & 0 & 0\\
c_1 c_3 & 0 & 0
\end{pmatrix},
\qquad &
\Lambda^{12} &= \frac{1}{2}
\begin{pmatrix}
- c_1 c_2 & 0 & 0\\
0 & c_1 c_2 & 0\\
c_2 c_3 & c_1 c_3 & 0
\end{pmatrix},
\\[3mm]
\Lambda^{22} &=
\begin{pmatrix}
0 & -c_1 c_2 & 0\\
0 & 0 & 0\\
0 & c_2 c_3 & 0
\end{pmatrix},
\qquad &
\Lambda^{13} &= \frac{1}{2}
\begin{pmatrix}
- c_1 c_3 & 0 & 0\\
- c_2 c_3 & 0 & c_1 c_2\\
0 & 0 & c_1 c_3
\end{pmatrix},
\\[3mm]
\Lambda^{33} &=
\begin{pmatrix}
0 & 0 & -c_1 c_3\\
0 & 0 & -c_2 c_3\\
0 & 0 & 0
\end{pmatrix},
\qquad &
\Lambda^{23} &= \frac{1}{2}
\begin{pmatrix}
0 & -c_1 c_3 & -c_1 c_2\\
0 & -c_2 c_3 & 0\\
0 & 0 & c_2 c_3
\end{pmatrix}.
\end{aligned}
\end{equation}
These satisfy the following identities
\begin{equation}\label{eq:LambdaIIOmega}
\Lambda^{11} + \Lambda^{22} + \Lambda^{33} = - \Omega,
\end{equation}
and
\begin{equation}
\Lambda^{IJ} \Omega + \Omega (\Lambda^{IJ})^T =0.
\end{equation}
We can define the following bispinors $(Z, \tilde{Z})$ that will enter the computation later
\begin{equation}\label{eq:AdS4bispinors}
Z^{IJ} = \zeta^I \otimes \bar{\zeta}^J , \qquad 
\tilde{Z}^{IJ} = \hat{\gamma}^{(4)} \zeta^I \otimes \bar{\zeta}^J ,
\end{equation}
These obey
\begin{equation}
d Z^{IJ}_{-} = m(\deg)\, Z^{IJ}_{+} , \qquad 
d \tilde{Z}^{IJ}_{+} = - m(\deg)\, \tilde{Z}^{IJ}_{-} ,
\end{equation}
where, given a $k$-form $C_k$, one has $(\deg) C_k = k C_k$. In terms of these one can also show
\begin{equation}
\iota_{k^{IJ}} \, \mathrm{vol}(\mathrm{AdS}_4) = -4 i \tilde{Z}^{IJ}_3 \,, \qquad 
\iota_{u^{IJ}} \, \mathrm{vol}(\mathrm{AdS}_4) = 4 Z^{IJ}_3, \qquad 
d k^{IJ} = 8 m \, Z^{IJ}_2 \,.
\end{equation}
We observe that
\begin{equation}
Z_0^{IJ}, \qquad Z_3^{IJ}, \qquad Z_4^{IJ}, \qquad \tilde{Z}_0^{IJ}, \qquad \tilde{Z}_1^{IJ}, \qquad \tilde{Z}_4^{IJ},
\end{equation}
are antisymmetric in $I \leftrightarrow J$, while
\begin{equation}
Z_1^{IJ}, \qquad Z_2^{IJ}, \qquad \tilde{Z}_2^{IJ}, \qquad \tilde{Z}_3^{IJ},
\end{equation}
are symmetric.

\subsection{Deriving seven dimensional conditions}\label{app:7dNeq3}

To derive the seven dimensional conditions, we first introduce the following bilinears defined on M$_7$
\begin{equation}
X^{IJ}_{n} = \frac{1}{n !} \chi^{I \dagger} \gamma_{a_1 ... a_n} \chi^{J} e^{a_1 ... a_n}, \qquad Y^{IJ}_{n} = \frac{1}{n !} \chi^{I c \dagger} \gamma_{a_1 ... a_n} \chi^{J} e^{a_1 ... a_n},
\end{equation}
where $\chi^{I} = \{\chi^1, \chi^2, \chi^3 \}$. We note that
\begin{equation}\label{eq:symmXIJYIJ}
\begin{split}
& \text{Re}X_0^{IJ}, \qquad \text{Re} Y_0^{IJ}, \qquad \text{Im} Y_0^{IJ}, \qquad \text{Re} X_1^{IJ}, \\[2mm]
& \text{Im} X_2^{IJ}, \qquad \text{Re} Y_3^{IJ}, \qquad \text{Im} Y_3^{IJ}, \qquad \text{Re}X_4^{IJ},
\end{split}
\end{equation}
are symmetric in $I \leftrightarrow J $, while
\begin{equation}\label{eq:antisymmXIJYIJ}
\begin{split}
& \text{Im} X_0^{IJ}, \qquad \text{Re} Y_1^{IJ}, \qquad \text{Im} Y_1^{IJ}, \qquad \text{Im} X_1^{IJ}, \qquad \text{Re} X_2^{IJ}, \\[2mm]
& \text{Im} Y_2^{IJ}, \qquad \text{Im} X_4^{IJ}, \qquad \text{Re} X_5^{IJ}, \qquad \text{Im} X_5^{IJ}, 
\end{split}
\end{equation}
are antisymmetric. The eleven dimensional spinors can be written in terms of the Killing spinor on AdS$_4$ $\zeta^I_{+}$, and the spinor on M$_7$ $\chi^I$
\begin{equation}\label{eq:app11dspinor}
\epsilon = \zeta^I_{+} \otimes \chi^I + \text{m.c.}.
\end{equation}
The one-form dual to the Killing vector is defined in \eqref{eq:bilinears11D}. Using \eqref{eq:app11dspinor}, we can decompose \eqref{eq:bilinears11D} in terms of bilinears defined on AdS$_4$ and M$_7$. It then takes the following form
\begin{equation}
K = e^{A} ( k^{IJ} \text{Re} X^{IJ}_0 - u^{IJ}\text{Im} X^{IJ}_0) - \Omega^{IJ} \text{Re} Y^{IJ}_1 + f^{IJ} \text{Im} Y^{IJ}_1 .
\end{equation}
Following our $4 + 7$ split \eqref{eq:Ansatz11dbosonic}, we find the affine connection decomposes as
\begin{equation}
\Gamma^{\mu}_{\nu \rho} =  \Gamma^{\mu (\text{AdS}_4)}_{\nu \rho}, \qquad \Gamma^{a}_{b c} =  \Gamma^{a (\text{M}_7)}_{b c}, \qquad \Gamma^{\mu}_{a \nu} = \partial_a A \, \delta^{\mu}_{\nu} \qquad \Gamma^{a}_{\mu \nu} = e^{2 A} g^{ab} \partial_b A \, g_{\mu \nu}^{\text{AdS}_4},
\end{equation}
while $\Gamma^{\mu}_{ab} = \Gamma^{a}_{\mu b} = 0 $, where with greek letters we refer to AdS$_4$ directions and with latin letters to M$_7$. Now we substitute these in \eqref{eq:BPS11} to end up with constraints involving only data on M$_7$. We begin with \eqref{eq:BPS11a}, from $\nabla_{(a} K_{b)} = 0$ we find
\begin{equation}
\begin{split}
& \nabla^{(7)}_{(a} \text{Re} (Y^{IJ}_1)_{b)} = 0, \qquad  \nabla^{(7)}_{(a} \text{Im} (Y^{IJ}_1)_{b)} = 0.
\end{split}
\end{equation}
$\nabla_{(\mu} K_{a)} = 0$ yields
\begin{equation}
\begin{split}
& d (e^{-A} \text{Im} X^{IJ}_0 ) = m e^{- 2 A} \text{Im} Y^{IJ}_1 , \qquad d (e^{-A} \text{Re} X^{IJ}_0 ) =0. 
\end{split}
\end{equation}
$\nabla_{(\mu} K_{\nu)} = 0$ leads to
\begin{equation}
\begin{split}
& {\cal{L}}_{\text{Im} Y^{IJ}_1} A = - m e^{-A} \text{Im} X^{IJ}_0, \qquad {\cal{L}}_{\text{Re} Y^{IJ}_1} A = 0.
\end{split}
\end{equation}
In summary, \eqref{eq:BPS11a} provides the following set of constraints
\begin{subequations}
\begin{align}
\nabla^{(7)}_{(a} \text{Re} (Y^{IJ}_1)_{b)} &= 0,
& \qquad \mathcal{L}_{\text{Re} Y^{IJ}_1 } A &= 0,
& \qquad d \bigl(e^{-A} \text{Re} X_0^{IJ} \bigr) &= 0, \label{eq:noAdS51}\\[2mm]
\nabla^{(7)}_{(a} \text{Im} (Y^{IJ}_1)_{b) } &= 0 ,
& \qquad \mathcal{L}_{\text{Im} Y^{IJ}_1} A &= - m e^{-A} \text{Im} X_0^{IJ},
& \qquad d \bigl(e^{-A} \text{Im} X_0^{IJ} \bigr) &= m e^{- 2 A} \text{Im} Y^{IJ}_1, \label{eq:noAdS52}
\end{align}
\end{subequations}
Looking at \eqref{eq:noAdS52}, one can show that in general we have an enhancement to AdS$_5$ unless we impose 
\begin{equation}
\text{Im} Y^{IJ}_1 = \text{Im} X_0^{IJ} = 0. 
\end{equation}
This is the same argument which appeared in \cite{Conti:2024rqy,Macpherson:2021lbr,Legramandi:2020txf}, we omit the details and refer to these references for more details. Focusing on \eqref{eq:noAdS51}, these imply $\text{Re} (Y^{IJ}_{1})_{b}$ are Killing vectors with respect to the whole eleven dimensional geometry and also that
\begin{equation}
\text{Re} X_0^{IJ} = e^A c^{IJ},
\end{equation}
where $c^{IJ}$ are constants. However under the assumption that no component of $\chi^I$ vanishes, we can exploit a GL(3,$\mathbb{R}$) symmetry to reabsorb these constants into the definitions of $\zeta^I$, so that, without loss of generality we can take
\begin{equation}
\text{Re} X_0^{IJ} = e^A \delta^{IJ}.
\end{equation}
In this way, the one-form becomes
\begin{equation}\label{eq:11dKilling}
K = e^{2A} k - \text{Re} Y_1,
\end{equation}
where we defined $ k = \delta^{IJ} k^{IJ}$ and $\text{Re} Y_1 = \Omega^{IJ} \text{Re} Y^{IJ}_1$. The dual Killing vector is
\begin{equation}
K^{A} \partial_{A} = k^{\mu} \partial_{\mu} - \text{Re} Y_1^a \partial_a .
\end{equation}
Similarly to the $4+7$ split happening for the one-form, we proceed in the same way for the two- and five-form\footnote{The 4d and 7d matrices share the same symmetries in their indices.}
\begin{align}
\Xi & = - 4 e^{2A} \left( Z_2^{IJ} \text{Re} Y_0^{IJ} + i \tilde{Z}_2^{IJ} \text{Im} Y_0^{IJ} \right) + 4 e^A \left( Z_1^{IJ} \wedge \text{Re} X_1^{IJ} - i \tilde{Z}_1^{IJ} \wedge \text{Im} X_1^{IJ} \right) \nn \\[2mm]
& + \Omega^{IJ} \text{Re} X_2^{IJ} + f^{IJ} \text{Im} Y_2^{IJ}, \\[2mm]
\Sigma & = 4 e^{4A} \left( - Z_4^{IJ} \wedge \text{Re} Y_1^{IJ} + i  \tilde{Z}_4^{IJ} \wedge \text{Im} X_1^{IJ} \right) - 4 e^{3 A} \left( - Z_3^{IJ} \wedge \text{Re} X_2^{IJ} + i \tilde{Z}_3^{IJ} \wedge \text{Im} X_2^{IJ} \right) \nn \\[2mm]
& - 4 e^{2 A} \left( Z_2^{IJ} \wedge \text{Re} Y_3^{IJ} + i \tilde{Z}_2^{IJ} \wedge \text{Im} Y_3^{IJ} \right) + 4 e^{A} \left( Z_1^{IJ} \wedge \text{Re} X_4^{IJ} - i \tilde{Z}_1^{IJ} \wedge \text{Im} X_4^{IJ} \right) \nn \\[2mm]
& - \Omega^{IJ} \text{Re} Y_5^{IJ} - f^{IJ} \text{Im} X_5^{IJ},
\end{align}
where $(Z_p^{IJ}, \tilde{Z}_p^{IJ})$ are the $p$-forms components of the polyform bilinears defined in \eqref{eq:AdS4bispinors}.

The Kosmann Lie derivative acting on $\epsilon$ is defined as \cite{Tomasiello:2022dwe}
\begin{equation}
{\cal{L}}_K \epsilon = K^A \nabla_A \epsilon + \frac{1}{4} \nabla_A K_B \Gamma^{AB} \epsilon.
\end{equation}
We then use
\begin{equation}\label{eq:AdS4covder}
\begin{aligned}
\nabla_{\mu}^{(11)} &= \nabla_{\mu}^{\text{AdS}_4} +\frac{1}{2} \Gamma_{\mu}dA , & \qquad & \mu={0,1,2,3}, \\[2mm]
\nabla_{a}^{(11)} &= \nabla_{a},& \qquad & a={4,..,9},
\end{aligned}
\end{equation}
to obtain
\begin{equation}
{\cal{L}}_K \epsilon = e^{2A} \delta^{IJ} ({\cal{L}}_{k^{IJ}} \zeta^K) \otimes \chi^K - \zeta^K \otimes (\Omega^{IJ} {\cal{L}}_{\text{Re} Y_1^{IJ}} \chi^K ) = 0.
\end{equation}
This combined with \eqref{eq:LiederivspinorAdS4} and \eqref{eq:LambdaIIOmega} implies
\begin{equation}
\Omega^{IJ} {\cal{L}}_{\text{Re} Y_1^{IJ}} \chi^K = - 2 m \Omega^{K L} \chi^L.
\end{equation}
We can factor out the $c_I$ and we end up with the equivalent constraint
\begin{equation}\label{eq:LiederivcalS}
{\cal{L}}_{\text{Re} Y_1^{IJ}} \chi^K = m ({\cal{S}}^{IJ})^{K L} \chi^L,
\end{equation}
where ${\cal{S}}^{IJ}$ are the SO($3$) generators consistent with the R-symmetry of ${\cal{N}}=3$ \eqref{eq:defcalS}. The spinors $\chi^I$ transform in the $\mathbf{3}$ of SO($3$). Consequently, the bilinears defined in \eqref{eq:XIJYIJ} transform as
\begin{equation}
\mathbf{3} \otimes \overline{\mathbf{3}} = \mathbf{1} \oplus \mathbf{3} \oplus \mathbf{5},
\end{equation}
where we identify $\mathbf{1}$ with the trace, $\mathbf{3}$ the antisymmetric and $\mathbf{5}$ the symmetric and traceless component of the general $ 3 \times 3$ matrices $X_n^{IJ}$ and $Y_n^{IJ}$. For convenience, we define the vector $\xi^K$ as the dual of the antisymmetric matrix $\text{Re}Y_1^{IJ}$
\begin{equation}
\text{Re}Y_1^{IJ} = \frac{1}{2} \epsilon^{IJK} \xi^{K},
\end{equation}
that can be also written as $\xi^{K} = \epsilon^{IJK} \text{Re}Y_1^{IJ}$. By means of the Levi Civita tensor properties, $ ({\cal{S}}^{IJ})^{K L} = - \epsilon^{M K L}\epsilon^{I J M}$. In this way we can rewrite \eqref{eq:LiederivcalS} as
\begin{equation}\label{eq:LiederivcalSfinal}
{\cal{L}}_{\xi^I} \chi^J = - 2 m \epsilon^{I J K} \chi^K.
\end{equation}
The Lie derivative, with respect to $K$, of each of the eleven dimensional forms, including $K$, is zero \cite{Gauntlett:2003wb}. We then find that this constraint on $K$ leads to
\begin{equation}
{\cal{L}}_K K = 2 e^{2A} \, k \, {\cal{L}}_{\text{Re} Y_1} A = 0,
\end{equation}
upon using \eqref{eq:noAdS52}. Plugging our $ 4 + 7 $ split in \eqref{eq:BPS11} we obtain the symmetric conditions
\begin{subequations}\label{eq:7dsusyallsymm}
\begin{align}
& d (e^{2 A} \text{Re} Y_0^{IJ}) - 2 m e^A \text{Re} X_1^{IJ} = 0, \\[2mm]
& 3 m e^{2 A} \text{Im}  Y_0^{IJ} - c \delta^{IJ} = 0, \\[2mm]
& d (e^{2A} \text{Im} Y_0^{IJ} ) = 0, \\[2mm]
& d (e^A \text{Re} X_1^{IJ}) = 0, \\[2mm]
& 6 e^A \delta^{IJ} \star_7 dA - \text{Im} X_2^{IJ} \wedge G_4 = 0, \\[2mm]
& (6 e^{4A} m \delta^{IJ} - 2 c \text{Im} Y_0^{IJ} ) \text{vol}_7 - e^{4A} \text{Im} Y_3^{IJ} \wedge G_4 = 0, \\[2mm]
& 2 c \text{Re} Y_0^{IJ} \text{vol}_7 + e^{4A} \text{Re} Y_3^{IJ} \wedge G_4 = 0, \\[2mm]
& d (e^{3A} \text{Im} X_2^{IJ} )+ 3 m e^{2 A} \text{Im} Y_3^{IJ}- e^{4A} \delta^{IJ} \star_7 G_4 = 0, \\[2mm]
& d (e^{2A} \text{Re} Y_3^{IJ} ) - 2 m e^A \text{Re} X_4^{IJ} + e^{2A} \text{Re} Y_0^{IJ} G_4 = 0, \\[2mm]
& d (e^{2A} \text{Im} Y_3^{IJ} ) + e^{2A} \text{Im} Y_0^{IJ} G_4 = 0, \\[2mm]
& d (e^A \text{Re} X_4^{IJ} ) - e^A \text{Re} X_1^{IJ} \wedge G_4 = 0,
\end{align}
\end{subequations}
and the antisymmetric conditions
\begin{subequations}\label{eq:7dsusyallantisymm}
\begin{align}
& d (e^A \text{Im} X_1^{IJ}) + m \text{Im}Y_2^{IJ} = 0, \\[2mm]
& d (\text{Re} Y_2^{IJ}) - i_{\text{Re}Y_1^{IJ}} G_4 = 0, \\[2mm]
& d (\text{Im} Y_2^{IJ}) = 0, \\[2mm]
& 3 e^{4A} \star_7 d \text{Re} Y_1^{IJ} + c \text{Re} Y_5^{IJ} - 2 e^{4A} \text{Re} Y_2^{IJ} \wedge \star_7 G_4 - e^{4A} \text{Im} Y_1^{IJ} \wedge G_4 = 0, \\[2mm]
& 2 e^{4A} \text{Im} Y_2^{IJ} \wedge \star_7 G_4 - e^{4A} \text{Re} Y_1^{IJ} \wedge G_4 - c \text{Im} Y_5^{IJ} = 0, \\[2mm]
& \text{Re}X_2^{IJ} \wedge G_4 = 0, \\[2mm]
& d \left( e^{4A} \text{Re} Y_1^{IJ} \right) - 4 m e^{3A} \text{Re}X_2^{IJ} + c \text{Im} Y_2^{IJ} = 0, \\[2mm]
& d ( e^{4A} \text{Im} Y_1^{IJ} ) + e^{4A} i_{\text{Re} Y_1^{IJ}} \star_7 G_4 - c \text{Re} Y_2^{IJ} = 0, \\[2mm]
& d (e^{3A} \text{Re} X_2^{IJ} ) = 0, \\[2mm]
& d (e^A \text{Im} X_4^{IJ}) + m \text{Im} Y_5^{IJ} - e^A \text{Im} X_1^{IJ} \wedge G_4 = 0, \\[2mm]
& d(\text{Re} Y_5^{IJ}) + c e^{- 4 A} i_{\text{Re} Y_1^{IJ} } \text{vol}_7 + \text{Re} Y_2^{IJ} \wedge G_4 = 0, \\[2mm]
& d(\text{Im} Y_5^{IJ}) + \text{Im} Y_2^{IJ}\wedge G_4 = 0.
\end{align}
\end{subequations}
We observe that the diagonal part contains 3 copies of the necessary and sufficient conditions for ${\cal{N}}=1$ AdS$_4$ \eqref{eq:AdS4N=1}. It thus follows that necessary and sufficient conditions for supersymmetry are given by the diagonal part of \eqref{eq:7dsusyallsymm} provided that each component of $\chi^I$ is independent.

\section{More details on the ${\cal{N}}=2$ derivation} \label{sec:AppendixNeq2}

This section provides further details on some of the steps involved in recovering the ${\cal{N}}=2$ AdS$_4$ class of \cite{Gabella:2012rc}.

We start by giving more details on the explicit forms of the vielbein $E_i$. As we outlined above, since there is a U($1$) isometry in our metric, we align $E_1$ along it and we fix the overall function by imposing it has unit norm. Regarding the other two vielbein directions, we can derive them by just imposing that they are orthogonal among themselves and $E_1$, unit norm, and such that
\begin{equation}
E_i E_i = e^{i + 4} e^{i + 4}.
\end{equation}
We report them in the $(\alpha, \beta)$ parametrisation
\begin{align}\label{eq:Eiei}
E_1 & = - \frac{3 e^{3A} m \cos \beta e^7 + (3 e^{3A} m \cos^2 \alpha + c \sin^2 \alpha) \sin \beta e^6 + (c - 3 e^{3A} m ) \cos \alpha \sin \alpha \sin \beta e^5 }{\sqrt{9 e^{6A} m^2 (1 - \sin^2 \alpha \sin^2 \beta) +c^2 \sin^2 \alpha \sin^2 \beta}}, \nn \\[2mm]
E_2 & = \frac{ ( \cos \alpha \sin \beta  e^7 + \cos \beta (\sin \alpha e^5 - \cos \alpha e^ 6 )) }{\sqrt{ 1 - \sin^2 \alpha \sin^2 \beta }}, \\[2mm]
E_3 & = - \frac{1}{\sqrt{1 - \sin^2 \alpha \sin^2 \beta}\sqrt{9 e^{6A} m^2 (1 - \sin^2 \alpha \sin^2 \beta ) + c^2 \sin^2 \alpha \sin^2 \beta }} \nn \\[2mm]
& ( \cos \alpha (3 e^{3A} m + (c - 3 e^{3A} m) \sin^2 \alpha \sin^2 \beta ) e^5 + \sin \alpha (3 e^{3A} m - (c \cos^2 \alpha + 3 e^{3A} m \sin^2 \alpha) \sin^2 \beta) e^6 \nn \\[2mm]
& - c \sin \alpha \sin \beta \cos \beta e^7 ). \nn
\end{align}

\section{A Canonical frame for the d=7 Majorana spinor} \label{app:canonical frame}

For a seven dimensional unit norm Majorana spinor in Euclidean dimension, we can define the following bilinears
\begin{equation}\label{eq:canonphi3}
- \frac{i}{3!} \eta^{\dagger} \gamma_{a b c} \eta e^{abc} = \phi_3 , \qquad - \frac{1}{4!} \eta^{\dagger} \gamma_{a b c d} \eta e^{abcd} = \star_7 \phi_3,
\end{equation}
where we work in conventions such that $i \gamma_{1234567} = 1$. These can be used to define also a seven-form through the following $\text{vol}_7 = \frac{1}{7} \phi_3 \wedge \star_7 \phi_3 $. The three- and four-form can be decomposed in terms of SU(3)-structure forms as
\begin{equation}
\phi_3 = J \wedge V - \text{Im}\Omega, \qquad \star_7 \phi_3 = \frac{1}{2} J \wedge J + \text{Re}\Omega \wedge V.
\end{equation}
For a single seven dimensional Majorana spinor in Euclidean dimension, there always exists a canonical frame \cite{Macpherson:2024qfi} such that
\begin{equation}\label{eq:frame7dMajorana}
\gamma_{1234} \eta = \gamma_{1256} \eta = \gamma_{1357} \eta = - \eta.
\end{equation}
In this frame, the SU(3)-structure forms are
\begin{equation}
J = e^{12} + e^{34} + e^{56}, \qquad \Omega = (e^1 + i e^2) \wedge (e^3 + i e^4) \wedge (e^5 + i e^6), \qquad V = e^7.
\end{equation}
One is always free to align, without loss of generality, a single arbitrary Majorana spinor $\hat{\eta}$ along $\hat{\eta} = || \hat{\eta} || \eta $ through a rotation of the vielbein on M$_7$. Any further additional Majorana spinor, $\tilde{\eta}$, can also be decomposed in the basis of $\eta $ through two real arbitrary functions $(a,b)$ and a unit norm one-form $V$ as $\tilde{\eta} = || \tilde{\eta} || (a \eta -i b  V \eta)$ such that $a^2+b^2=1$, look at \eqref{eq:M7decspinor} for example. An arbitrary number of spinors can be expanded in this basis, but each comes with a distinct unit norm one-form and pair of functions \cite{Dibitetto:2018ftj}.

\subsection{SU(3)-structure in terms of SU(2)}\label{subapp:SU3SU2}

In section \ref{sec:Neq2class} we find it useful to decompose the SU(3)-structure in terms of an SU(2)-structure. This can be achieved as follows.

In this canonical basis we define
\begin{align}\label{eq:tildej1j2}
i_{e^6} \Omega & = - (\tilde{j}_1 + i \tilde{j}_2), \qquad i_{e^5} \Omega = i (\tilde{j}_1 + i \tilde{j}_2),
\end{align}
while we write the K\"ahler form as
\begin{equation}\label{eq:tildej3}
J = j_3 + e^5 \wedge e^6.
\end{equation}
In this way, the 7d metric can be written as
\begin{equation}
ds^2(\text{M}_7) = ds^2(\text{SU}(2)) + e^{i+4} e^{i+4} = ds^2(\text{SU}(2)) + E_i^2.
\end{equation}
where $E_i$ are given in \eqref{eq:Eiei}. In subsection \ref{subsec:TriSasaki} we will make use of the $\tilde{j}_i$ introduced in \eqref{eq:tildej1j2}-\eqref{eq:tildej3}. However, for the ${\cal{N}}=2$ class in Section \ref{sec:Neq2class}, it is more convenient to introduce the following rotated basis of two-forms
\begin{equation}
j_1 = \tilde{j}_1, \qquad j_2 = \cos \beta \tilde{j}_2 - \sin \beta \tilde{j}_3, \qquad j_3 = \cos \beta \tilde{j}_3 + \sin \beta \tilde{j}_2.
\end{equation}
Notice that both satisfy
\begin{equation}
j_i \wedge j_j = \tilde{j}_i \wedge \tilde{j}_j = 2 \delta_{ij} e^1 \wedge e^2 \wedge e^3 \wedge e^4.
\end{equation}

\printbibliography

\end{document}